# RECURRENT ALGORITHMS FOR DETECTION OF STOCHASTIC SIGNALS IN THE STATE SPACE

Arthur N. Yuryev

30th Central Scientific Research Institute, Ministry of Defence (USSR)

**Abstract**

The paper is devoted to synthesis of recurrent algorithms for detection of stochastic signals given in state space. The structure of the algorithms synthesized is shown to be close to that of the Kalman filter. Analysis of one of the algorithms synthesized is carried out. Illustration of connection between weight coefficients of processing system, which are formed in implicit form, is given. Dynamics of amplification and feedback coefficients of the system is studied; calculation of its detection characteristics is fulfilled. Synthesis of the filters is also carried out for the continuous time. The algorithms synthesized are extended to the case of a mixture of stochastic correlated interferences and white noise. Modification of one of the algorithms is made which enables its use for solving problems of multialternative detection.[1]

[1] For the Russian text see: Y. A. Furman, A. N. Yuryev and V. V. Yanshin, *Numerical methods of binary image processing and recognition*. Krasnoyarsk University, Krasnoyarsk, Russia, 1992; ISBN 5-7470-0204-X.

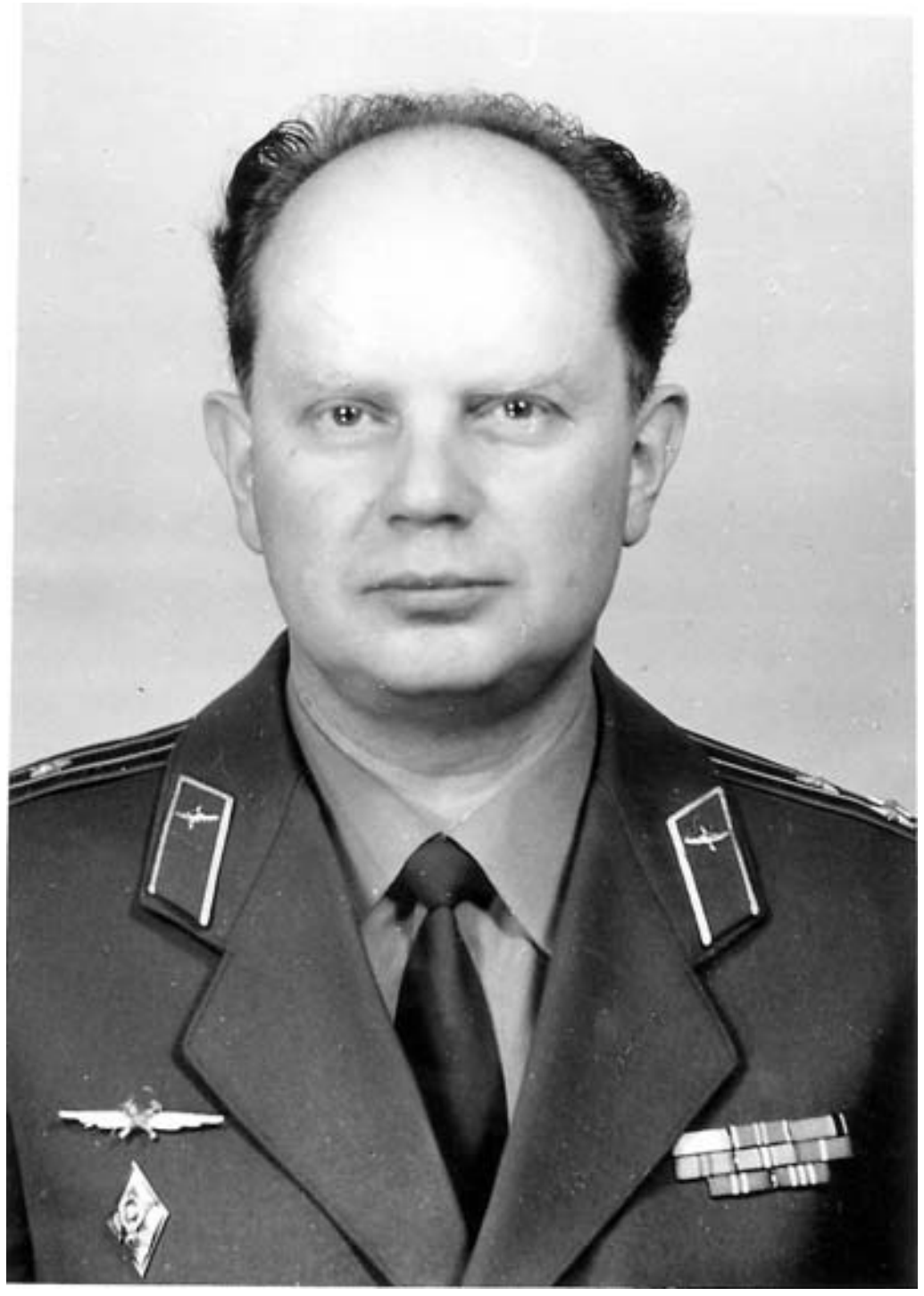

**Arthur Nikolaevich YURYEV**

A. N. Yuryev was born at Rybinsk city, Russia, on 5 July 1932. He graduated in 1956 with honors and engineer degree (Masters Degree) in radio engineering and received the candidate of sciences degree (PhD) in 1962 from Prof. N. E. Zhukovsky Air Force Engineering Academy, Moscow, USSR. Later, in 1976, he received the Doctor of Sciences degree. The ranks of senior research scientist and professor in radar and radio navigation were conferred to him in 1965 and 1984, respectively.

From 1950 he was with the USSR Air Force and from 1962 he worked at the 30th Central Scientific Research Institute of the Ministry of Defense of USSR also known as the Central Research Institute of Aviation and Space Technology.

The field of his scientific interests included, inter alia, radar and navigation, image recognition, information theory, communication and control theory and their applications. He was an author (with coauthors) of a number of monographs such as “Correlation-Extremum Methods of Navigation” (Moscow, 1982), “Adaptive Control Systems of Aircrafts” (Moscow, 1987), “Numerical methods of binary image processing and recognition” (Krasnoyarsk, 1992).

Col. A. N. Yuryev was a member of the A. S. Popov All-Union Society of Radio, Electronics and Communication Engineering.

Prof. A. N. Yuryev deceased on September 30, 1990.

# RECURRENT ALGORITHMS FOR DETECTION OF STOCHASTIC SIGNALS IN THE STATE SPACE

A. N. Yuryev

*Abstract.* This paper is devoted to synthesis of recurrent algorithms for detection of stochastic signals given in state space.

The structure of algorithms synthesized is shown to be close to that of Kalman filter.

Analysis of one of the algorithms synthesized is carried out. Illustration of connexion between weight coefficients of processing system, which are formed in implicit form, is given. Dynamics of amplification and feedback coefficients of the system is studied; calculation of its detection characteristics is fulfilled.

Synthesis of the filters is also carried out for continuous time. The algorithms synthesized are spread to the case of mixture of stochastic correlated interferences and white noise. Modification of one of the algorithms is made, which allows its use for solving problems of multialternative detection.[1]

---

[1] Professor. A.N.Yuryev prematurely deceased on September 30, 1990. This parer, initially written by A.N.Yuryev in Russian as a book capter on the basis of a number of journal articles published in Russian journals, was prepared for publication in English by Dr. V.A.Yuryev in 1991. Some textual editing was made and some additional material was added to the present article in comparison with the published book chapter. Present address for contacts concerning this paper: Dr. V.A.Yuryev, Prokhorov General Physics Institute of the Russian Academy of Sciences, 38, Vavilov street, Moscow, 119991, Russia. ( V.Y.).

## INTRODUCTION.

There exist a wide bibliography on theory and methods of stochastic signal detection [1-7]. In the paper proposed to your attention synthesis and analysis of new algorithms for detection of stochastic signals is carried out. Their peculiarities are as follows: the algorithms are of recursive character, they functionate in a state space, they are direct analogs of Kalman filters for detection problems. These algorithms allows to simplify a detection procedure in comparison with known ones [1-7].

The algorithms obtained are able to solve detection problems in various phase spaces, in particular in multi-dimensional phase spaces defined with that of parameters of signal or object movement which are subsequently subjected to filtering and estimating. Such parameters might be an object velocity and acceleration, parameters of its trajectory, etc. Signal detection in such phase spaces may have certain advantages in comparison with traditional techniques of signal detection owing to using of information about object trajectory and its movement character.

For one of the algorithms synthesized the analysis is made in the paper. It is shown that possibility of the algorithm presentation in compact recurrent form is conditioned by characteristic properties of its weight factor matrices and determined either with used principle of current (implicit) formation of these matrices or with Markov model of random process. For stationary input process amplification and feedback factors of the system are stabilized rapidly when the algorithm operate (in two-three steps of observation). So far as the detection efficiency is concerned, the algorithm considered, being much simpler in its structure, practically does not differ from the standard optimal one.

The algorithms obtained for discrete time and uncorrelated observation noises are spread to the cases of a continuous time and signal detection in mixture of correlated stochastic interferences and uncorrelated clutter.

And at last the synthesis of algorithm for multialternative detection is made.

# 1. TWO-ALTERNATIVE DETECTION.

## 1.1. DISCRETE TIME. ALGORITHM SYNTHESIS.

### 1.1.1. Problem statement.

Let us consider that $n_o$-dimensional observation vector $\mathbf{z}(i)$ given in discrete time i, $i=\overline{1,n}$, may be formed in accordance with two hypothesis: $\eta_o$ (signal is absent) and $\eta_1$ (signal is present).

$$\begin{aligned} &\eta_o:\ \mathbf{z}(i)=\mathbf{n}(i) \\ &\eta_1:\ \mathbf{z}(i)=\mathbf{H}(i)\mathbf{x}(i)+\mathbf{n}(i) \end{aligned} \qquad (1.1.1)$$

In equation (1.1.1) $\mathbf{x}(i)$ is $m_o$-dimensional signal vector; $\mathbf{n}(i)$ is $n_o$-dimensional noise vector; $\mathbf{H}(i)$ is matrix of dimension $n_o \times m_o$ which converts signal vector to the space of observed vectors. Suppose $\mathbf{X}(i)$ and $\mathbf{n}(i)$ to be normally distributed, independent, zero-mean random vectors. Suppose also interference to be a weighed sum of K independent and uncorrelated in time interfering signals $\mathbf{n}_k(i)$, $k=\overline{1,K}$, and internal noise $\mathbf{n}_o(i)$, i.e.

$$\mathbf{n}(i)=\sum_{k=1}^{K}\mathbf{A}_k(i)\mathbf{n}_k(i) \qquad (1.1.2)$$

where $\mathbf{n}_k(i)$ are $n_k$-dimensional vectors, $\mathbf{A}_k(i)$ are matrices of dimension $n_o\times n_k$.

The covariance matrices of the observed process $\mathbf{z}(i)$ are:

$$\begin{aligned}&\eta_o:\ \mathbf{K}_o=[\mathbf{K}_o(i,j)];\\&\mathbf{K}_o(i,j)=E[\mathbf{n}(i)\mathbf{n}^T(j)]=\mathbf{N}(i)\delta_{ij};\\&\eta_1:\ \mathbf{K}_1=[\mathbf{K}_1(i,j)];\\&\mathbf{K}_1(i,j)=\mathbf{K}_s(i,j)+\mathbf{K}_o(i,j)=\mathbf{H}(i)\mathbf{R}_s(i,j)\mathbf{H}^T(j)+\mathbf{N}(i)\delta_{ij};\end{aligned} \qquad (1.1.3)$$

where $\mathbf{R}_s(i,j)=E[\mathbf{x}(i)\mathbf{x}^T(j)]$; $\delta_{ij}$ is the Kronecher symbol; E[...] means a symbol of statistical average; superscript T is a sign of transponating.

The matrices $\mathbf{K}_o(i,j)$ and $\mathbf{K}_1(i,j)$ represent block square symmetric ones with a number of blocks $n\times n$ in each of them and with a number of elements in the blocks $\mathbf{K}_o(i,j)$ and $\mathbf{K}_1(i,j)$ equal to $n_o\times n_o$.

Let us assume that linear Markov model is correct for the signal, i.e. the signal is a solution of linear difference equation:

$$\mathbf{x}(i)=\mathbf{S}(i,i-1)\mathbf{x}(i-1)+\mathbf{A}(i-1)\mathbf{w}(i-1) \qquad (1.1.4)$$

where $\mathbf{S}(i,i-1)$ is $m_o\times m_o$-dimensional matrix; $\mathbf{w}(i-1)$ is $l_o$-dimensional vector of discrete uncorrelated normal noise; $\mathbf{G}(i-1)$ is a matrix of dimension $m_o\times l_o$; $E[\mathbf{w}(i)]=\mathbf{0}$; $E[\mathbf{n}(i)\mathbf{w}^T(j)]=0$; $E[\mathbf{x}(i)\mathbf{w}^T(j)]=0$; $i\leqslant j$; $i,j=\overline{1,n}$.

The problem is to find the signal processing algorithm which discriminates between the hypothesis $\eta_o$ and $\eta_1$ in the best way.

Notice that when solving the problem stated one has to use two ways of description of the signal $\mathbf{x}(i)$ statistical properties: using the covariance matrices $\mathbf{R}_s(i,j)$ and by use of (1.1.4). The well-known relation connecting these two ways is [6,7]:

$$\mathbf{R}_s(i,j)=\begin{cases}\mathbf{S}(i,j)\mathbf{R}_s(j,j), & i\geqslant j;\\ \mathbf{R}_s(i,i)\mathbf{S}^T(j,i), & j\geqslant i.\end{cases}$$

1.1.2. **Synthesis.**

The sufficient statistics, which is the output effect of the optimal processing system, may be expressed as

$$y_o=\sum_{i=1}^{n}\sum_{j=1}^{n}\mathbf{z}^T(i)\mathbf{W}_o(i,j)\mathbf{z}(j) \qquad (1.1.5)$$

where

$$\mathbf{W}_o(i,j)=\mathbf{M}_o(i,j)-\mathbf{L}_o(i,j) \qquad (1.1.6)$$

matrices $\mathbf{M}_o(i,j)$ and $\mathbf{L}_o(i,j)$ are $n_o\times n_o$-blocks of inverse covariance matrices: $\mathbf{K}_o^{-1}=[\mathbf{M}_o(i,j)]$, $\mathbf{K}_1^{-1}=[\mathbf{L}_o(i,j)]$.

Let us consider two following signal processing algorithms based on (1.1.5):

*a. Processing algorithm of the form:*

$$\mathbf{y}_1(n)=\sum_{i=1}^{n}\mathbf{z}^T(i)\mathbf{U}_o(i) \text{ or } \mathbf{y}_1(i)=\mathbf{y}_1(i-1)+\mathbf{z}^T(i)\mathbf{U}_o(i),$$

$$y_1(0)=0,\ i=\overline{1,n} \qquad (1.1.7)$$

where

$$\mathbf{U}_o(i)=\sum_{j=1}^{i}\mathbf{W}(i,j)\mathbf{z}(i) \qquad (1.1.8)$$

is an output signal of linear part of the processing system; matrix $\mathbf{W}(i,j)$ may be considered as discrete pulse transient response of the system linear part. Note that $\mathbf{W}(i,j)=\mathbf{W}^T(j,i)$.

*b. Processing algorithm using the following presentation of the matrix* $\mathbf{W}(i,j)$ [8,9]:

$$\mathbf{W}(i,j)=\sum_{k=1}^{n} \mathbf{b}(i,k)\mathbf{b}^T(j,k) \qquad (1.1.9)$$

Presentation (1.1.9) is valid for every symmetric matrix $\mathbf{W}=[\mathbf{W}(i,j)]$. The algorithm considered may be expressed in the following equations:

$$y_2(n)=\sum_{k=1}^{n} \mathbf{U}^T(k)\mathbf{U}(k) \text{ or } y_2(k)=y_2(k-1)+\mathbf{U}^T(k)\mathbf{U}(k),$$

$$y_2(0)=0,\ k=\overline{1,n} \qquad (1.1.10)$$

where

$$\mathbf{U}(k)=\sum_{j=1}^{k} \mathbf{b}^T(j,k)\mathbf{z}(j). \qquad (1.1.11)$$

Presentation of the matrix (1.1.6) in the form (1.1.9) does not determine unambiguously the form of the matrix $\mathbf{b}(i,k)$. To eliminate this ambiguity let us consider that the matrix $\mathbf{b}(i,k)$ like the matrix $\mathbf{W}(i,j)$ satisfies the property

$$\mathbf{b}(i,k)=\mathbf{b}^T(k,i). \qquad (1.1.12)$$

It should be pointed out that the equation (1.1.12) allows to reduce the signal processing algorithm (1.1.10) to recursive form. Taking (1.1.12) into account, the equations (1.1.9) and (1.1.11) may be transformed into:

$$\mathbf{W}(i,j)=\sum_{k=1}^{n}\mathbf{b}(i,k)\mathbf{b}(k,j), \qquad (1.1.13)$$

$$\mathbf{U}(k)=\sum_{j=1}^{n}\mathbf{b}(k,j)\mathbf{z}(j), \qquad (1.1.14)$$

the matrix $\mathbf{b}(k,j)$ may be considered as discrete puls transient response of the processing system (1.1.10), (1.1.14) linear part.

Below we shall make a synthesis of detection algorithms using signal presentation in the state space (1.1.4). Let us start consideration from the algorithm (1.1.10), (1.1.14). Results concerning the algorithm (1.1.7), (1.1.8) will be obtained as a consequence of the algorithm (1.1.10), (1.1.14) consideration.

Transform now the equation (1.1.13) for $\mathbf{b}(i,k)$ matrix. Multiplying both parts of the equation (1.1.13) by $\mathbf{K}_1(j,m)$ from the right and by $\mathbf{K}_O(l,i)$ from the left and summarizing obtained equations by i and j and allowing for

$$\mathbf{W}(i,j)=\mathbf{M}(i,j)-\mathbf{L}(i,j),$$

$$\sum_{i=1}^{l}\mathbf{K}_O(l,i)\mathbf{M}(i,j)=\mathbf{I}\delta_{jl}, \qquad (1.1.15)$$

$$\sum_{j=1}^{l}\mathbf{L}(i,j)\mathbf{K}_1(j,m)=\mathbf{I}\delta_{im}, \quad i,m\leqslant l,$$

($\mathbf{I}$ is a unit matrix) we have:

$$\sum_{k=1}^{n}\Big[\sum_{i=1}^{l}\mathbf{K}_O(l,i)\mathbf{b}(i,k)\sum_{j=1}^{l}\mathbf{b}(k,j)\mathbf{K}_1(i,m)\Big]=\mathbf{K}_s(l,m). \qquad (1.1.16)$$

As the observation noise $\mathbf{n}(i)$ is white, equation (1.1.16) may be presented in the form:

$$\sum_{k=1}^{n} \{\mathbf{N}(l)\mathbf{b}(l,k)[\mathbf{b}(k,m)\mathbf{N}(m)+\sum_{j=1}^{l} \mathbf{b}(k,j)\mathbf{K}_s(j,m)]\}=\mathbf{K}_s(l,m). \quad (1.1.17)$$

As it is shown in [10] equation (1.1.17) is equivalent to difference equation

$$\sum_{k=1}^{n} \{[\mathbf{D}(l)\mathbf{b}(l,k)-\mathbf{S}_o(l,l-1)\mathbf{b}(l-1,k)][\mathbf{b}(k,m)\mathbf{N}(m)+ \quad (1.1.18)$$
$$+\sum_{j=1}^{l} \mathbf{b}(k,j)\mathbf{K}_s(j,m)]=0$$

where

$$\mathbf{D}(l)=\mathbf{N}^{-1}(l)\mathbf{L}^{-1}(l,l) \quad (1.1.19)$$

$$\mathbf{S}_o(l,l-1)=\mathbf{N}^{-1}(l)\mathbf{H}(l)\mathbf{S}(l,l-1)\mathbf{H}^{-1}(l-1)\mathbf{N}(l-1). \quad (1.1.20)$$

When deriving equations (1.1.19) and (1.1.20), we assumed that the inverse matrices composing them exist; if $\mathbf{H}(l-1)$ is not a square matrix, $\mathbf{H}^{-1}(l-1)$ should be considered as quasi-inverse one [11,12]. Furthermore it was assumed that proper values of the matrix $\mathbf{Q}(l)=\mathbf{W}(l,l)\mathbf{N}(l)=\mathbf{I}-\mathbf{L}(l,l)\mathbf{N}(l)$, $l=\overline{1,n}$, $\lambda_p$ satisfy the condition $|\lambda_p|<1$, $p=\overline{1,n_o}$. Estimate of this requirement implementation is made in [13].

Let us consider the equation (1.1.18). For the matrix $\mathbf{K}_1(j,m)=\mathbf{K}_s(j,m)+\mathbf{N}(j)\delta_{im}$ is positively determined, expression in square brackets in (1.1.18) cannot equal zero when $\mathbf{b}(k,m)$ does not identically equal zero. Hence the equation (1.1.18) have an untrivial solution only when the condition

$$\mathbf{D}(l)\mathbf{b}(l,k)=\mathbf{S}_o(l,l-1)\mathbf{b}(l-1,k)$$

is realized, therefore

$$\mathbf{b}(l,k)=\mathbf{F}(l,l-1)\mathbf{b}(l-1,k) \qquad (1.1.21)$$

where

$$\mathbf{F}(l,l-1)=\mathbf{D}^{-1}(l)\mathbf{S}_{o}(l,l-1)= \\ =\mathbf{L}(l,l)\mathbf{H}(l)\mathbf{S}(l,l-1)\mathbf{H}^{-1}(l-1)\mathbf{N}(l-1). \qquad (1.1.22)$$

The equation (1.1.21) defines the system transient response matrix $\mathbf{b}(l,k)$. From the relation (1.1.21) a connection between the matrix $\mathbf{W}(i,j)$, which determines the detection algorithm (1.1.8) and (1.1.9), and parameters describing a state space may be found out. Multiplying from the right both parts of the equation (1.1.21) by $\mathbf{b}(k,m)$ and summarizing over k, taking account of (1.1.13), we have:

$$\begin{cases} \mathbf{W}(l,m)=\mathbf{F}(l,l-1)\mathbf{W}(l-1,l), \\ \mathbf{W}(m,l)=\mathbf{W}(m,l-1)\mathbf{F}^{T}(l,l-1), \quad m=\overline{1,l-1}. \end{cases} \qquad (1.1.23)$$

So the equations for transient response matrices of linear parts of the algorithms (1.1.7), (1.1.8) and (1.1.10), (1.1.14) are identic.

Recur to the expression (1.1.14) linear part of the signal processing system (1.1.10). Regarding (1.1.21) this expression may be rewritten as follows:

$$\mathbf{U}(l)=\sum_{k=1}^{l}\mathbf{b}(l,k)\mathbf{z}(k)=\sum_{k=1}^{l-1}\mathbf{b}(l,k)\mathbf{z}(k)+\mathbf{b}(l)\mathbf{z}(l)=$$

$$=\mathbf{F}(l,l-1)\sum_{k=1}^{l-1}\mathbf{b}(l-1,k)\mathbf{z}(k)+\mathbf{b}(l)\mathbf{z}(l), \quad \mathbf{b}(l)=\mathbf{b}(l,l).$$

Then we have:

$$\mathbf{U}(l)=\mathbf{F}(l,l-1)\mathbf{U}(l-1)+\mathbf{b}(l)\mathbf{z}(l) \qquad (1.1.24)$$

The formula (1.1.24) is a recursive form of functioning algorithm of the processing system linear part; the parameter $\mathbf{b}(l)=\mathbf{b}(l,l)$ is an amplification factor of the system linear part.

The structure of recurrent filter is determined by the relations (1.1.10) and (1.1.24) to which it is necessary to add the start condition for difference equation (1.1.24): $\mathbf{U}(0)=0$.

As far as the processing algorithm (1.1.8), (1.1.9) is concerned, the subsequent recursive equation is:

$$\mathbf{U}(l)=\mathbf{F}(l,l-1)\mathbf{U}(l-1)+\mathbf{W}(l)\mathbf{z}(l);\ \mathbf{U}(0)=0;\ \mathbf{W}(l)=\mathbf{W}(l,l) \qquad (1.1.25)$$

1.1.3. **Summary.**

So the decision of the problem stated has been obtained. The recurrent detection algorithms have been synthesized, which form the output signals of detector ($y_1(l)$ and $y_2(l)$) in the following way:

$$\begin{cases} y_1(l)=y_1(l-1)+\mathbf{z}^T(l)\mathbf{U}_o(l);\ y_1(0)=0;\ l=\overline{1,n}; \\ \mathbf{U}_o(l)=\mathbf{F}(l.l-1)\mathbf{U}_o(l-1)+\mathbf{W}(l)\mathbf{z}(l);\ \mathbf{U}_o(0)=\mathbf{0}. \end{cases} \qquad (1.1.26)$$

$$\begin{cases} y_2(l)=y_2(l-1)+\mathbf{U}^T(l)\mathbf{U}(l);\ y_2(0)=0;\ l=\overline{1,n}; \\ \mathbf{U}(l)=\mathbf{F}(l.l-1)\mathbf{U}(l-1)+\mathbf{b}(l)\mathbf{z}(l);\ \mathbf{U}(0)=\mathbf{0}. \end{cases} \qquad (1.1.27)$$

The output signals $y_1(l)$ and $y_2(l)$ are formed for every value of current time l; then the values of $y_1(n)$ or $y_2(n)$ are compared with fixed threshold, and as a result the decision about truth of $\eta_o$ or $\eta_1$ hypothesis is made.

The structural scheme of the algorithms are shown in Fig.1.1.1 (*a* is the algorithm (1.1.26), *b* is the algorithm

(1.1.27)), block $\gamma_{del}=1$ means the one-time-delay device, block 1 means a shift register, block 2 is a threshold device.

The matrix $\mathbf{F}(l,l-1)$, which plays the part of the processing systemfeedback factor, is determined by the expression (1.1.22).

The matrix factor $\mathbf{W}(l)$ in (1.1.26) is the matrix $\mathbf{W}(i,j)$ calculated for current time l (i=j=l) and acting as an amplification coefficient of the system based on (1.1.26).

$$\mathbf{W}(i,j)=\mathbf{N}^{-1}(i)\delta_{ij}-\mathbf{L}(i,j); \tag{1.1.28}$$

the matrix $\mathbf{L}(i,j)$ is calculated from the equation

$$\sum_{j=1}^{l} \mathbf{L}(i,j)\mathbf{K}_1(j,m)=\mathbf{I}\delta_{im};\ i,m\leqslant l, \tag{1.1.29}$$

where $\mathbf{I}$ is a unit matrix of dimension $n_o\times n_o$.The equation (1.1.29) expresses an operation of calculation of matrix, which is inverse with respect to left upper submatrix of the block matrix $\mathbf{K}_1$ limited by current time $l\leqslant n$. let us designate this matrix as $\mathbf{K}_{1l}$. The matrix $\mathbf{W}(l,j)=\mathbf{W}^T(j,l)$ satisfies the equation (1.1.23) written for l-th row and l-th column of a block matrix

$$\mathbf{W}_l=[\mathbf{W}_l(i,j)]=\begin{cases}[\mathbf{W}(l,j)],\ j\leqslant l;\\ [\mathbf{W}(i,l)],\ i\leqslant l,\ l=\overline{1,n}.\end{cases} \tag{1.1.30}$$

The matrix amplification factor of the system based on (1.1.27) $\mathbf{b}(l)$ is the matrix $\mathbf{b}(i,j)=\mathbf{b}^T(j,i)$ (i=j=l) connected with the matrix $\mathbf{W}_l(i,j)$ by the equation (1.1.13). The following relations written for l-th row and l-th column are valid for the matrix $\mathbf{b}=[\mathbf{b}(i,j)]$:

$$\begin{cases}\mathbf{b}(l,k)=\mathbf{F}(l,l-1)\mathbf{b}(l-1,k);\\ \mathbf{b}(k,l)=\mathbf{b}(k,l-1)\mathbf{F}(l,l-1),\ k=\overline{1,n}\end{cases} \tag{1.1.31}$$

In conclusion we shall emphasize the difference between the relations (1.1.23) and (1.1.31). The matrix $\mathbf{b}=[\mathbf{b}(i,j)]$ is degenerated; the equations (1.1.31) allow obtaining every row (column) if the matrix coefficients $\mathbf{F}(l,l-1)$, $l=\overline{2,n}$, are known. The equation (1.1.23) connects only a part of the matrix $\mathbf{W}_l$ row, which consists of a diagonal element and elements laying on the left of it, with laying below part of the next row (corresponding parts of columns are also connected). These properties of the matrices $\mathbf{W}_l$ and $\mathbf{b}$, stipulated by current character of the matrix $\mathbf{W}_l$ formation and Markov signal model, determine a recursive form of the algorithms (1.1.26) and (1.1.27).

### 1.1.4. **Brief discussion.**

As it is seen from the schemes in Fig 1.1.1 the structure of linear parts of the synthesized detection algorithms (which are in the dashed boxes in Figs.1.1.1, a, b) is close to Kalman filter (the scheme of Kalman filter forming the estimate $\mathbf{x}^*(l)$ of input signal $\mathbf{z}(l)$ is given for comparison in Fig.1.1.1, c). Distinction lies in absence of observed signal "prediction" circuit ($\mathbf{H}(l)$-block) and in different content of amplification blocks $\mathbf{W}(l)$, $\mathbf{b}(l)$ and $\mathbf{K}(l)$ and signal transformation blocks $\mathbf{F}(l,l-1)$ and $\mathbf{A}(l,l-1)$. Structural closeness of the obtained algorithms and Kalman filter assures a possibility of their realization on the common technological basis and make possible their mutual conversion by changing corresponding software. This circumstance allows using common system elements either for detection or for posterior filtering of the signal.

## 1.2. ANALYSIS.

### 1.2.1. Illustration of (1.1.23) relation implementation. Dynamics of the algorithm (1.1.26) coefficients.

Let us make an analysis of the algorithm (1.1.26) in conformity with stationary Markov signal model with parameter $S(i,j)=r$ and corresponding covariance function $K_s(i,j)=R_s(i,j)=\sigma_s^2 r^{|i-j|}$ where $\sigma_s^2$ is signal power (assume $H(i)=1,\ i=\overline{1,n}$). We shall consider that the noise is also stationary, i.e. $N(i)=\sigma_n^2=const$. The matrix $\mathbf{W}_l=[\mathbf{W}_l(i,j)]$ for this situation, when $l=4$, is:

$$
\mathbf{W}_l=\left[\begin{array}{ccccccc}
\frac{1}{pd_1} & \xrightarrow{\mathcal{F}(2,1)} & \frac{r}{d_2} & \xrightarrow{\mathcal{F}(3,2)} & \frac{r^2}{d_3}p & \xrightarrow{\mathcal{F}(4,3)} & \frac{r^3}{d_4}p^2 \\
\Big\downarrow{\scriptstyle \mathcal{F}(2,1)} & & & & & & \\
\frac{r}{d_2} & & \frac{1}{\sigma_n^2}-\frac{1}{\sigma_s^2}\Phi\frac{d_1}{d_2} & \xrightarrow{\mathcal{F}(3,2)} & \frac{r}{d_3}(1+p-r^2) & \xrightarrow{\mathcal{F}(4,3)} & \frac{r^2}{d_4}p(1+p-r^2) \\
\Big\downarrow{\scriptstyle \mathcal{F}(3,2)} & & \Big\downarrow{\scriptstyle \mathcal{F}(3,2)} & & & & \\
\frac{r^2}{d_3}p & & \frac{r}{d_3}(1+p-r^2) & & \frac{1}{\sigma_n^2}-\frac{1}{\sigma_s^2}\Phi\frac{d_2}{d_3} & \xrightarrow{\mathcal{F}(4,3)} & \frac{r}{d_4}\{(1+p)^2-r^2[2+p-r^2(1-p)]\} \\
\Big\downarrow{\scriptstyle \mathcal{F}(4,3)} & & \Big\downarrow{\scriptstyle \mathcal{F}(4,3)} & & \Big\downarrow{\scriptstyle \mathcal{F}(4,3)} & & \\
\frac{r^3}{d_4}p^2 & & \frac{r^2}{d_4}p(1+p-r^2) & & \frac{r}{d_4}\{(1+p)^2-r^2[2+p-r^2(1-p)]\} & & \frac{1}{\sigma_n^2}-\frac{1}{\sigma_s^2}\Phi\frac{d_3}{d_4}
\end{array}\right]
$$

(1.2.1)

Notice that in the algorithm (1.1.26) only the diagonal terms $\mathbf{W}_l(l)=\mathbf{W}(l)$ of the matrix $\mathbf{W}_l$ are present in explicit form, the rest elements of the matrix $\mathbf{W}_l$ are formed implicitly by using the relations (1.1.23).

The following designations are accepted in the matrix (1.2.1):

$d_1=\sigma_s^2\Delta_1$; $\Delta_1=1+p$; $p=\sigma_n^2/\sigma_s^2$;

$d_2=\sigma_s^2\Delta_2$; $\Delta_2=(1+p)^2-r^2$;

$d_3=\sigma_s^2\Delta_3$; $\Delta_3=(1+p)^3-2r^2(1+p)+r^4(1-p)$;

$$d_4=\sigma_s^2\Delta_4;\ \Delta_4=[(1+p)^4-r^4]-r^2(1+p)(1+p-r^2)(1+r^2)-$$
$$-r^4p[1+p+r^2(1+p)].$$

The parameters $\Delta_l$, $l=\overline{1,4}$, comprise determinants of the $\sigma_s^{-2}\mathbf{K}_{ll}$ matrices. The parts of matrix (1.2.1) rows, which are constituted by diagonal element and elements laying on the left of it, are connected with elements of the next row laying below by the F(l,l-1), $l=\overline{2,4}$, factors; the same may be said about the columns of the matrix $\mathbf{W}_l$. These connections are shown in (1.2.1) as arrows. So the example given illustrates the validity of the relations (1.1.23) obtained as a result of the system synthesis. The coefficients F(l,l-1) have been calculated for the matrix (1.2.1) by using the formula (1.1.22); they may be expressed in terms of the determinants $\Delta_l$:

$$F(2,1)=rp\frac{\Delta_1}{\Delta_2};\ F(3,2)=rp\frac{\Delta_2}{\Delta_3};\ F(4,3)=rp\frac{\Delta_3}{\Delta_4}.$$

In accordance with the mathematical induction method

$$F(l,l-1)=rp\frac{\Delta_{l-1}}{\Delta_l};$$

Diagonal elements of the matrix (1.2.1) make up corresponding amplification factors W(l) of the system using the algorithm (1.1.26). They may be expressed as:

$$W(l)=\frac{1}{\sigma_n^2}\left[1-p\frac{\Delta_{l-1}}{\Delta_l}\right].$$

For comparison with (1.2.1) we shall adduce the matrix $\mathbf{W}_n=[\mathbf{W}_n(i,j)]$ which is analogous to the matrix $\mathbf{W}_l=[\mathbf{W}_l(i,j)]$ but obtained when the matrix $\mathbf{K}_l$ is being inverted for given n. Elements of the matrix $\mathbf{W}_n$ refer to weight factors of well known classical optimal algorithm for stochastic signal detection. An output signal of the system realizing this algorithm is

$$y_o=\sum_{i=1}^{n}\sum_{j=1}^{n}\mathbf{z}^T(i)\mathbf{W}_n(i,j)\mathbf{z}(j) \qquad (1.2.2)$$

where $\mathbf{W}_n(i,j)=\mathbf{N}^{-1}(i,j)\delta_{ij}-\mathbf{L}_n(i,j)$ and the matrix $\mathbf{L}_n$ is $n_o\times n_o$-block of the inverse covariance matrix $\mathbf{K}_n^{-1}=[\mathbf{L}_n(i,j)]$. For the same initial data, for which the matrix (1.2.1) is obtained, the matrix $\mathbf{W}_n$ (n=l=4) is:

$$\mathbf{W}_n=\begin{bmatrix} \frac{1}{\sigma_n^2}-\frac{d_3}{\sigma_s^2 d_4} & \frac{r}{d_4}\{(1+p)^2-r^2[2+p-r^2(1-p)]\} & \frac{r^2}{d_4}p(1+p-r^2) & \frac{r^3}{d_4}p^2 \\ \frac{r}{d_4}\{(1+p)^2-r^2[2+p-r^2(1-p)]\} & \frac{1}{\sigma_n^2}-\frac{d_3}{\sigma_s^2 d_4} & \frac{r}{d_4}[(1+p)^2-r^2](1+p-r^2) & \frac{r^2}{d_4}p(1+p-r^2) \\ \frac{r^2}{d_4}p(1+p-r^2) & \frac{r}{d_4}[(1+p)^2-r^2](1+p-r^2) & \frac{1}{\sigma_n^2}-\frac{d_3}{\sigma_s^2 d_4} & \frac{r}{d_4}\{(1+p)^2-r^2[2+p-r^2(1-p)]\} \\ \frac{r^3}{d_4}p^2 & \frac{r^2}{d_4}p(1+p-r^2) & \frac{r}{d_4}\{(1+p)^2-r^2[2+p-r^2(1-p)]\} & \frac{1}{\sigma_n^2}-\frac{d_3}{\sigma_s^2 d_4} \end{bmatrix} \qquad (1.2.3)$$

Dwell on the main distinctions between the $\mathbf{W}_l$ and $\mathbf{W}_n$ matrices. The matrix $\mathbf{W}_n$ is renewed entirely when n changes whereas with every new step l the new (l-th) row and column are simply added to the matrix $\mathbf{W}_l$, if dimensions of the $\mathbf{W}_l$ and $\mathbf{W}_l$ matrices being equal their last row and column coinciding. Thus the relation (1.1.23) provides current formation of the matrix $\mathbf{W}_l$ while the number of observations increases. Noteworthy also that the matrix $\mathbf{W}_n$, being a part of the algorithm (1.2.2) structure, does not allow simplifications in processing algorithm like that which take place when using the $\mathbf{W}_l$ matrix.

Let us consider now the amplification and feedback factors W(l) and F(l,l-1) for the conditions accepted in this section. The results of working out of these factors are given in Fig.1.2.1 versus the current time l. It results from the given dependencies that at stationary input processes the factors W(l) and F(l,l-1) stabilize rather rapidly (at the second or third observation step); only at high signal correlation stabilization is slightly dragged out. The settled value of the amplification coefficient reduces while the value of F(l,l-1) on the contrary rise when signal correlation increases. When signal-to-noise ratio decreases the value of the amplification factor falls and the feedback factor rises. At hard signal correlation, when coherent detection can be provided, the dependency of these coefficients on p abruptly weakens with l increase, and the role of the feedback factor increasing ($F(l,l-1)\rightarrow 1$ when $l\rightarrow\alpha$) and the importance of a new information decreasing ($W(l)\rightarrow 1$ when $l\rightarrow\alpha$), that is quite reasonable from physical viewpoint. If the signal correlation is utterly absent the feedback in the algorithm linear part vanishes (F(l,l-1)=0) and signal accumulation is accomplished only in incoherent way; it is in accordance with conventional view either.

Analysis of the algorithm dynamics in the case of stationary input processes shows that at given signal and noise characteristics only two practically unchanged parameters are used in the algorithm (1.1.26). They are the feedback and amplification factors. While the algorithm (1.2.2) requires knowing of n(n+1)/2 weight coefficients. This circumstance bears witness that the algorithm (1.1.26) is considerably simpler in its structure then the algorithm (1.2.2).

For more details see also [13,14].

### 1.2.2. **Detection characteristics.**

To calculate the detection characteristics we shall lean upon the results given in [15]. If the output signal of the system may be presented as a quadratic form

$$y=\mathbf{Z}^{T}\mathbf{V}\mathbf{Z}, \qquad (1.2.4)$$

where $\mathbf{Z}=\{\mathbf{z}(1),\dots,\mathbf{z}(n)\}^{T}$ is normally distributed vector random process with average value equal to zero and covariance matrix **K**; **V** is a block square matrix with a number of blocks n×n, then the probability of exceeding of some threshold T by the random value y is determined by the following relation:

$$P = 1 - \frac{1}{\pi}\int_{0}^{\infty}\mathbf{Re}\left\{\frac{exp(-j\xi T)-1}{-j\xi}\;\frac{d\xi}{det\left[\mathbf{I}-2j\xi\mathbf{K}\mathbf{V}\right]}\right\}. \qquad (1.2.5)$$

If the process considered is described by complex model then instead of vector **Z** in (1.2.4) and instead of **K** and **V** matrices in (1.2.5) one have to use the complex conjugate quantities.

The algorithm (1.1.26) considered here covers with regard for (1.1.23), except for the matrix $\mathbf{W}_1$ diagonal elements, also its elements placed at the left of the diagonal. Therefore this algorithm is equivalent to the algorithm (1.2.4) provided $\mathbf{V}=(1/2)(\mathbf{W}_1+\mathbf{W}'_1)$, where $\mathbf{W}'_1$ is a diagonal matrix whose diagonal coincides with that of $\mathbf{W}_1$. When estimating the efficiency of the the optimal algorithm (1.2.2), $\mathbf{V}=\mathbf{W}_n$. When considering both algorithms, to calculate the false alarm probability one has to assume $\mathbf{K}=\mathbf{K}_o$, and to calculate the true detection probability it is necessary to assume $\mathbf{K}=\mathbf{K}_1$.

For the process model accepted above, the computations of the detection characteristics of the system realizing the algorithm (1.1.26) has been made using the formula (1.2.5). The results of these computations are plotted in Fig.1.2.2 as graphs of inverse detection probability logarithm vs signal-to-noise ratio $q=p^{-1}$. The computations made indicate that the systems based on the algorithm (1.1.26) and (1.2.2) practically do not differ by their detection efficiency. So e.g. when r=0.8 the algorithm (1.2.2) assures at high values of SNR only by several hundredths of percent higher detection probability than the algorithm (1.1.26). At moderate SNR values (about 7dB) the detection probability for (1.2.2) is by several tenths of percent higher than that for (1.1.26). These distinctions become less as the rate of the signal fluctuations increase. And for uncorrelated signal the distinctions vanish at all. The false alarm probability as usually was accepted $10^{-4}$.

1.2.3. **Example of space-time signal processing.**

The algorithm considered may be propagated to the case of space-time signal detection. Let us assume the signal receiving to be realized by discrete earial antenna system; the signals coming to the elements of earial system form the phase space, in which information processing is made. Formally the problem statement and solution given above are not changed but the form of matrices figuring in the relations (1.1.1), (1.1.2) and (1.1.4) are concretized. The matrices **H**(i) and $\mathbf{A}_k(i)$ are loosing the dependency on i and become diagonal; their sizes correspond to the number of discrete elements of the antenna $n_o$:

$$\mathbf{H}(i)=\mathbf{H}=diag[exp\{\tilde{j}(2\pi/\lambda)\mathbf{r}_1\rho_s\},...,exp\{\tilde{j}(2\pi/\lambda)\mathbf{r}_n\rho_s\}];$$

$$\mathbf{A}_k(i)=\mathbf{A}_k=diag[exp\{\tilde{j}(2\pi/\lambda)\mathbf{r}_1\rho_k\},...,exp\{\tilde{j}(2\pi/\lambda)\mathbf{r}_{n_o}\rho_k\}];$$

here $\tilde{j}$ is a complex unit; $\lambda$ is a wavelength; **r** is a vector-radius of p-th earial element center, $p=\overline{1,n_o}$; $\rho_s$, $\rho_k$ are unit vector-radii directed to the sources of signal and k-th interference respectively, $k=\overline{1,K}$; $\mathbf{A}_o(i)=\mathbf{I}$.

The matrices **S**(i,i-1) and **R**(i-1) also become diagonal with the sizes $n_o \times n_o$ and as a rule are expressed via the unit matrix.

The calculations of directivity characteristics of the space-time processing system, built on equidistant linear antenna array, has been made on the basis of the results obtained. The directivity characteristics of such a system has been determined as a dependency of its output signal y(l) for the current moment of time on the generalized angular parameter $\varphi=(2\pi\Delta/\lambda)sin\gamma$ when the signal

$$\mathbf{z}(l)=\mathbf{z}=[exp\{\tilde{j}((n_o+1)/2-1)\varphi\},...,exp\{\tilde{j}((n_o+1)/2-n_o)\varphi\}]^+$$

is inputting the system ($\Delta$ is the distance between the phase centers of the array elements, $\gamma$ is the angle count from the normal to the line of the phase centers).

The computation has been made for three-element array ($n_o$=3) with one source of active interference (K=1); it has been assumed that $\mathbf{r}_p\rho_s$=0 and $\mathbf{r}_p\rho_1=(\lambda/6)[(n_o+1)/2-p]$, i.e. the signal and the interference has been coming from the directions $\varphi$=0 and $\varphi_1=\pi/3$ respectively. The directivity characteristics has been computed using the formulae (1.1.26) with the signal and the interference covariance functions of the following type:

$$\mathbf{K}_s(i,j)=\sigma_s^2 exp\{-\alpha|i-j|\}[exp\{-\beta|p-s|\}];\ \mathbf{K}_o(i,j)=\mathbf{N}\delta_{ij};$$

$$\mathbf{N}=[\sigma_{n1}^2 exp\{-\gamma|p-s|+\tilde{j}(p-s)\pi/3\}+\sigma_{n0}^2\delta_{ps}];\ p,s=\overline{1,n_o}.$$

Under the above conditions

$$\mathbf{S}(l,l-1)=e^{-\alpha}\mathbf{I};\ \mathbf{F}(l,l-1)=e^{-\alpha}\mathbf{L}(l)\mathbf{N};\ \mathbf{W}(l)=\mathbf{N}^{-1}\mathbf{L}(l).$$

The following initial data have been accepted for the computations: $\alpha=10^{-2}$; $\beta=10^{-3}$; $\gamma=10^{-3}$. The results of the computations are given in Fig.1.2.3, where the output effect of the system $y_n(l)$ is normalized so that at $\varphi$=0, $y_n(l)$=1. The designations in Fig.1.2.3 correspond to the following meanings of energetic parameters: 1 is $\sigma_{n0}^2$=10, $\sigma_{n1}^2$=0; 2 is $\sigma_{n0}^2$=10, $\sigma_{n1}^2$=10; 3 is $\sigma_{n0}^2$=1, $\sigma_{n1}^2$=10; the directivity characteristics depends on $\sigma_s^2$ weakly. The dependency of normalized directivity characteristics on l is practically absent, so the results adduced are true for every l. It follows from Fig.1.2.3 that the minimum reception area is formed in the interference direction and the reception level is depen-dent on the ratio between interference and noise.

## 1.3. CONTINUOUS TIME.

Let us consider the modification of the algorithms synthesized in the Section 1.1 for the case of a continuous time [16].

### 1.3.1. **Problem statement.**

Let us assume that n-dimensional observation vector $\mathbf{z}(t)$, given in continuous time t ($0<t<T$), may be formed in accordance with two hypotheses: $\eta_0$ (signal is absent) and $\eta_1$ (signal is present):

$$\begin{aligned} &\eta_0:\ \mathbf{z}(t)=\mathbf{n}(t);\\ &\eta_1:\ \mathbf{z}(t)=\mathbf{H}(t)\mathbf{x}(t)+\mathbf{n}(t); \end{aligned} \qquad (1.3.1)$$

where $\mathbf{x}(t)$ is n-dimensional signal vector; $\mathbf{n}(t)$ is n-dimensional noise vector; $\mathbf{H}(t)$ is a matrix of dimension n×m; $\mathbf{x}(t)$ and $\mathbf{n}(t)$ are Gaussian independent random process with mathematical expectation equal to zero. The covariance matrix of observed process $\mathbf{z}(t)$ look like follows:

$$\begin{aligned} &\eta_0:\ \mathbf{K}_i(t,t_1)=E[\mathbf{n}(t)\mathbf{n}^+(t_1)]=\mathbf{N}_i(t)\delta(t-t_1);\\ &\eta_1:\ \mathbf{K}_{si}(t,t_1)=\mathbf{K}_s(t,t_1)+\mathbf{K}_i(t,t_1)=\\ &\qquad\qquad =\mathbf{H}(t)\mathbf{R}_s(t,t_1)\mathbf{H}^+(t_1)+\mathbf{N}_i(t)\delta(t-t_1); \end{aligned} \qquad (1.3.2)$$

where $\mathbf{R}_s(t,t_1)=E[\mathbf{x}(t)\mathbf{x}^+(t_1)]$; $\delta(t-t_1)$ is a Dirac delta function; superscript "+" is a symbol of Hermitian conjugation; E[...] means a statistical average. Assume that signal satisfies the following differential equation:

$$\dot{\mathbf{x}}(t)=\mathbf{F}(t)\mathbf{x}(t)+\mathbf{G}(t)\mathbf{w}(t), \qquad (1.3.3)$$

where $\mathbf{F}(t)$ is a matrix of dimension m×m, $\mathbf{w}(t)$ is l-dimensional vector of Gaussian white noise, $\mathbf{G}(t)$ is a matrix of dimension m×l: $E[\mathbf{w}(t)]=\mathbf{0}$; $E[\mathbf{n}(t)\mathbf{w}^+(t_1)]=0$; $E[\mathbf{x}(t)\mathbf{w}^+(t_1)]=0$, $t < t_1$. The point over $\mathbf{x}(t)$ means operation of differentiation.

Let us find the signal processing algorithm which discriminates between the hypotheses $\eta_0$ and $\eta_1$ in the best way.

1.3.2. **Synthesis.**

Let us choose two following types of algorithms for random signal detection, which are considered in the detection theory [1,2,4], as a basis of statistical synthesis:

*a. Processing algorithm*

$$y_1(T)=\int_0^T \mathbf{z}^+(t)\mathbf{U}_0(t)dt, \tag{1.3.4}$$

where

$$\mathbf{U}_0(t)=\int_0^t \mathbf{W}(t,t_1)\mathbf{z}(t_1)dt_1$$

is output signal of processing system linear part; the matrix $\mathbf{W}(t,t_1)$ may be considered as a pulse transient response of the system linear part; $\mathbf{W}(t,t_1)=\mathbf{W}^+(t_1,t)$.

*b. Processing algorithm*

$$y_2(T)=\int_0^T \mathbf{U}^+(t)\mathbf{U}(t)dt, \tag{1.3.5}$$

using representation of $\mathbf{W}(t,t_1)$ matrix in the form of

$$\mathbf{W}(t,t_1)=\int_0^T \mathbf{b}(t,t_2)\mathbf{b}^+(t_1,t_2)dt_2. \tag{1.3.6}$$

In (1.3.5)

$$\mathbf{U}(t)=\int_0^t \mathbf{b}^+(t_1,t)\mathbf{z}(t_1)dt_1.$$

Assuming $\mathbf{b}(t,t_1)=\mathbf{b}^+(t_1,t)$ we have

$$\mathbf{W}(t,t_1)=\int_0^T \mathbf{b}(t,t_2)\mathbf{b}(t_2,t_1)dt_2, \tag{1.3.7}$$

$$\mathbf{U}(t)=\int_0^t \mathbf{b}(t,t_1)\mathbf{z}(t_1)dt_1. \qquad (1.3.8)$$

The matrix $\mathbf{b}(t,t_1)$ may be considered as a pulse transient response of linear part (1.3.6) of the processing system (1.3.5), (1.3.8).

The synthesis of the detection algorithm, which uses signal presentation in the state space, will be started with consideration of the procedure specified by the expressions (1.3.5) and (1.3.8).

Transform the equation (1.3.7). Multiplying its both parts from the left by $\mathbf{K}_i(\tau,t)$ and from the right by $\mathbf{K}_{si}(t_1,\tau_1)$ and integrating over t and $t_1$ with regard to

$$\mathbf{W}(t,t_1)=\mathbf{M}(t,t_1)-\mathbf{L}(t,t_1),$$

$$\int_0^\tau \mathbf{K}(\tau,t)\mathbf{M}(t,t_1)dt=\mathbf{I}\ \delta(t_1-\tau), \qquad (1.3.9)$$

$$\int_0^\tau \mathbf{L}(t,t_1)\mathbf{K}_{si}(t_1,\tau_1)dt=\mathbf{I}\ \delta(t-\tau_1)$$

($\mathbf{I}$ is the unit matrix) we have

$$\int_0^T \left[\int_0^\tau \mathbf{K}_i(\tau,t)\mathbf{b}(t,t_2)dt \int_0^\tau \mathbf{b}(t_2,t_1)\mathbf{K}_{si}(t_1,\tau_1)dt_1\right]dt_2=\mathbf{K}_s(\tau,\tau_1). \qquad (1.3.10)$$

As the observation noise is white, this equation may be rewritten as

$$\int_0^T \left\{\mathbf{N}_i(\tau)\mathbf{b}(\tau,t_2)\left[\mathbf{b}(t_2,\tau_1)\mathbf{N}_i(\tau_1)+\int_0^\tau \mathbf{b}(t_2,t_1)\mathbf{K}_s(t_1,\tau_1)dt_1\right]\right\} dt_2=\mathbf{K}_s(\tau,\tau_1). \qquad (1.3.11)$$

To keep an analogy with the results of the discrete system synthesis (see the Section 1.1.2), consider that delta function is asymmetric assuming that

$$\int_0^\tau \delta(\tau-t)dt=1.$$

The equation (1.3.11) is equivalent to the following equation (see [16]):

$$\int_0^T \left[\mathbf{b}'(\tau,t_2)-\mathbf{F}_0(\tau)\mathbf{b}(\tau,t_2)\right]\left[\mathbf{b}(t_2,\tau_1)\mathbf{N}_i(\tau_1)+\int_0^\tau \mathbf{b}(t_2,t_1)\mathbf{K}_s(t_1,\tau_1)dt_1\right]dt_2=0 \qquad (1.3.12)$$

where $\mathbf{F}_0(\tau)=\bar{\varphi}(\tau)-\mathbf{Q}(\tau)$;

$$\bar{\varphi}(\tau)=\mathbf{N}_i^{-1}(\tau)\left[\left(\mathbf{H}'(\tau)+\mathbf{H}(\tau)\mathbf{F}(\tau)\right)\mathbf{H}^{-1}(\tau)\mathbf{N}_i(\tau)-\mathbf{N}_i'(\tau)\right]; \qquad (1.3.13)$$

$\mathbf{Q}(\tau)=\mathbf{W}(\tau)\mathbf{N}(\tau)$; $\mathbf{W}(\tau)=\mathbf{W}(\tau,\tau)$; the stroke means a derivative (for functions of two arguments it is with respect to the first one).

When deriving (1.3.12), it has been assumed, like in the Section 1.1.2, that $\mathbf{H}^{-1}(\tau)$ and $\mathbf{N}_i^{-1}(\tau)$ exist; if the matrices $\mathbf{H}(\tau)$ and $\mathbf{N}_i(\tau)$ are not square, $\mathbf{H}^{-1}(\tau)$ and $\mathbf{N}_i^{-1}(\tau)$ should be considered as quasi-inverse.

Give here another form of (1.3.13). Multiplying both parts of (1.3.13) by $\mathbf{x}^+(t_1)$ from the right, $t_1<t$, and making statistical averaging, we have

$$\mathbf{R}_s'(t,t_1)=\mathbf{F}(t)\mathbf{R}_s(t,t_1),$$

from whence

$$\mathbf{F}(t)=\mathbf{R}_s'(t,t_1)\mathbf{R}_s^{-1}(t,t_1). \qquad (1.3.14)$$

It follows from (1.3.1) and (1.3.14) that

$$\mathbf{H}(\tau)\mathbf{F}(\tau)\mathbf{H}^{-1}(\tau)=\mathbf{K}_s'(\tau,t_1)\mathbf{K}_s^{-1}(\tau,t_1)$$

and, hence,

$$\bar{\varphi}(\tau)=\mathbf{N}_i^{-1}(\tau)\Big[\mathbf{H}'(\tau)\Big(\mathbf{H}^{-1}(\tau)+ +\mathbf{K}_s'(\tau,t_1)\mathbf{K}_s^{-1}(\tau,t_1)\Big)\mathbf{N}_i'(\tau)-\mathbf{N}_i'(\tau)\Big]. \qquad (1.3.15)$$

Remark that the matrix $\bar{\Delta}(\tau)=-\mathbf{F}^{-1}(\tau)$ may be considered as a matrix analog of the process $\mathbf{x}(t)$ correlation time. In particular, for one-dimensional case, taking into account that the derivative from $R_s(t,t_1)$ is taken on the right slope of correlation function and tending $t\to t_1$, we obtain $F=-(1/\Delta)$, where $\Delta=\Delta(t)$ is the ratio of $R_s(t,t_1)$ to its derivative at $t_1=t+0$ (this ratio may be treated as x(t) process correlation time for current moment of time). The higher x(t) correlation, the closer F(t) to zero; when $F(t)\to-\infty$, the process x(t) approaches to white noise.

Let us consider the equation (1.3.12). When analyzing, account must be taken that the detection process may be finished at any moment of time $\tau$. So putting $T=\tau$ in (1.3.12) we come to recognize that a solution of the equation (1.3.12), which is true for every $\tau$, may be obtained only if its integrand equals zero. As the matrix $\mathbf{K}_{si}(t_1,\tau_1)=\mathbf{K}_s(t_1,\tau_1)+\mathbf{N}_i(t_1)\delta(t_1-\tau_1)$ is positive definite, so at $\mathbf{b}(t_2,t_1)$ not equal identically zero the expression in the second square brackets in (1.3.12) cannot be equal to zero. Thus (1.3.12) have an untrivial solution only if

$$\mathbf{b}'(\tau,t_2)=\mathbf{F}(\tau)\mathbf{b}(\tau,t_2). \qquad (1.3.16)$$

Analogous equation is valid for the algorithm (1.3.4)

either, in what one can convince himself multiplying from the right both parts of (1.3.16) by $\mathbf{b}(t_2,\tau_1)$ and integrating over $t_2$ taking account of (1.3.7):

$$\mathbf{W}'(\tau,\tau_1)=\mathbf{F}_0(\tau)\mathbf{W}(\tau,\tau_1). \qquad (1.3.17)$$

The equations (1.3.16) and (1.3.17) determine the matrices of pulse transient responses of the algorithm (1.3.5) and (1.3.4) linear parts.

Differentiating $\mathbf{U}(\tau)$ (1.3.8) we have regarding (1.3.16)

$$\mathbf{U}'(\tau)=\int_0^\tau \mathbf{b}'(\tau,t_1)\mathbf{z}(t_1)dt_1+\mathbf{b}(\tau)\mathbf{z}(\tau)=$$

$$=\mathbf{F}_0(\tau)\int_0^\tau \mathbf{b}(\tau,t_1)\mathbf{z}(t_1)dt_1+\mathbf{b}(\tau)\mathbf{z}(\tau).$$

where $\mathbf{b}(\tau)=\mathbf{b}(\tau,\tau)$, consequently

$$\mathbf{U}'(\tau)=\mathbf{F}_0(\tau)\mathbf{U}(\tau)+\mathbf{b}(\tau)\mathbf{z}(\tau);\ \mathbf{U}(0)=\mathbf{0}. \qquad (1.3.18)$$

As applied to (1.3.4) we have

$$\mathbf{U}_0'(\tau)=\mathbf{F}_0(\tau)\mathbf{U}_0(\tau)+\mathbf{W}(\tau)\mathbf{z}(\tau);\ \mathbf{U}_0(0)=\mathbf{0}. \qquad (1.3.19)$$

The equations (1.3.18) and (1.3.19) together with the equations, which express the process of output signal formation in current time, determine the structure of the detection filters being synthesized.

$$y_1'(\tau)=\mathbf{z}^+(\tau)\mathbf{U}_0(\tau),\ 0<\tau<T; \qquad (1.3.20)$$

$$y_2'(\tau)=\mathbf{U}^+(\tau)\mathbf{U}(\tau),\ \ 0<\tau<T. \qquad (1.3.21)$$

The structural schemes of the filters (1.3.20) ,(1.3.19) and (1.3.21), (1.3.18) are given in Fig.1.3.1,a and b respectively. The linear parts structure of the algorithms synthesized reflects the structure of the state equation (1.3.3) and it is close to the structure of the Kalman filter plotted as an examp-le in Fig.1.3.1,c. Distinction of the synthesized algorithm from a Kalman filter, which forms the estimate $\mathbf{x}^*(t)$ of a signal, is in absence of observed signal prediction circuit ($\mathbf{H}(\tau)$ block) and in different substance of amplification ($\mathbf{W}(\tau)$, $\mathbf{b}(\tau)$ and $\mathbf{K}(\tau)$) and feedback ($\mathbf{F}_0(\tau)$ and $\mathbf{F}(\tau)$) blocks.

### 1.3.3. **Discussion.**

It has been believed thus far that the elements of the matrices $\mathbf{W}(\tau,\tau_1)$ and $\mathbf{b}(\tau,\tau_1)$ are ordinary (smooth) functions and $\mathbf{W}(\tau)=\mathbf{W}(\tau,\tau)$ and $\mathbf{b}(\tau)=\mathbf{b}(\tau,\tau)$ exist. In reality the functions $\mathbf{W}(\tau,\tau_1)$ and $\mathbf{b}(\tau,\tau_1)$ are singular, because they contain delta functions, and consequently the meanings of $\mathbf{W}(\tau)$ and $\mathbf{b}(\tau)$ turn into infinity. It is conditioned by idealization accepted, which is connected with representation of noise correlation function as delta function (for this the noise energy is infinitely large) and with edge effects arisen when the second integral equation in (1.3.9) is solved. To impart physical sense to the results obtained it is necessary to "attribute" some finite width, equal to correlation time, to the noise correlation function. It is also required to neglect the edge effects in the second integral equation in (1.3.9) when solved; the last is possible if signal correlation time $\Delta_s \ll \tau$, what beginning from some $\tau$ usually asserts.

Let us give the estimate of amplification and feedback coefficients of the filters synthesized and make an analysis of dependence of these coefficients from signal and interference parameters using one-dimensional filters as an example. An esti-mate is based on pass in (1.3.9) from integrals to integral sums with discretization interval equal to correlation time $\Delta_i$ in the first equation and to correlation time of signal and noise mixture $\Delta_{si}$ in the second equation. Fulfilled this procedure we obtain:

$$M(\tau)=(N_i\Delta_i)^{-1};\ L(\tau)=(N_{si}\Delta_{si})^{-1}$$

where $\Delta_i=N_i/K_i(\tau,\tau)$; $\Delta_{si}=N_{si}/K_{si}(\tau,\tau)$; $N_{si}=N_s+N_i$; $N_s$ is average energetic spectral density of the signal in the limits of spectral width; $K_{si}(\tau,\tau)=N_s\Delta_s^{-1}+N_i\Delta_i^{-1}$; $\Delta_s=N_s/K_s(\tau,\tau)$. For the case considered $\varphi(\tau)=-\Delta_s^{-1}$. The designation of possible dependence of $N_i$, $N_s$, $\Delta_i$ and $\Delta_s$ values on $\tau$ is omitted in the relations presented.

The following relations for the coefficients $W(\tau)=W$ and $F_o(\tau)=F_o$ are valid in the accepted way of signal and noise description in case of steady-state noise ($N'(\tau)=0$)

$$WN_i\Delta_i=1-(1-q/\delta)/(1+q)^2; \tag{1.3.22}$$

$$F_o\Delta_i=-(1/\delta)[1+\delta-q(1+q\delta)/(1+q)^2] \tag{1.3.23}$$

where $q=N_s/N_i$; $\delta=\Delta_s/\Delta_i$.

The plots presenting the dependences (1.3.22) and (1.3.23) are given in Fig.1.3.2. Taking account of (1.3.7) one may show that coefficients $b(\tau)=b$ is easy to calculate from known W using the relation $b=[(1+1/q)N_i]^{1/2}W$.

The relations and curves obtained allow to make some inferences. In particular at $q\rightarrow\infty$ the coefficients of the filters approach to definite limits: $W\rightarrow 1/N_i\Delta_i$; $F\rightarrow -\Delta_s^{-1}$; $b\rightarrow WN^{1/2}$. Growth of $\delta$ meanings beyond $\delta=4$ practically does not result in changes of W (and b neither). The feedback factor $F_o$ is negative and with growing of $\delta$ rise also approaching in the limit to the value of $q^2/(1+q)^2-1$, which is close to zero at large enough q. Note that, as it follows from (1.3.18) and (1.3.19), the closer is $F_o$ to zero, the higher is a degree of correlation of the filter linear part output signals $U_o(\tau)$ and $U(\tau)$ and, consequently, the more efficient is signal linear processing. On the contrary, the large $F_o$ absolute value, the closer the processes $U_o(\tau)$ and $U(\tau)$ to white noise, and the efficiency of signal processing in the filters' linear parts falling. It is seen also from Fig.1.3.2 that either amplification factor or feedback one rise with q enhancement, that is conditioned by increase of coming information reliability.

### 1.3.4. **Summary.**

Thus, it should be concluded from the above that the structure of each of the filters synthesized is rather simple in comparison with familiar ones [1:6]; their linear parts are close to the structure of the continuous Kalman filter, but the prediction circuit is absent in them and amplification and feedback blocks have different algorithmic matter. Under steady-state input effects the feedback and amplification coefficients of the filters stabilize quickly approaching to definite limit meanings with rising of the ratio of signal correlation time to noise correlation time.

# 2. DETECTION AGAINST THE BACKGROUND OF CORRELATED INTERFERENCES AND NOISE.

The above algorithms have been synthesized under the condition of uncorrelateness of the noises and interferences in time, that considerably limits the area of their application. This restriction is lifted in this section; the results obtained in previous sections are shown to be a direct basis for solving the problem of stochastic signal detection in the mixture of correlated stochastic interferences and discrete white clutters [17], that greatly widen the sphere of practical application of the obtained algorithms.

## 2.1. PROBLEM FORMULATION.

Let as usually $n_o$-dimensional vector $\mathbf{z}(i)$ given in discrete time i, $i=\overline{1,n}$, to be formed in accordance with two hypotheses: $\eta_o$ (signal is absent) and $\eta_1$ (signal is present),

$$\begin{aligned} &\eta_o\text{: } \mathbf{z}(i)=\mathbf{H}_i(i)\mathbf{x}(i)+\mathbf{n}(i); \\ &\eta_1\text{: } \mathbf{z}(i)=\mathbf{H}_s(i)\mathbf{x}_s(i)+\mathbf{H}_i(i)\mathbf{x}_i(i)+\mathbf{n}(i). \end{aligned} \qquad (2.1.1)$$

In (2.1.1) $\mathbf{x}_s$ is $m_o$-dimensional signal vector; $\mathbf{x}_i$ is $k_o$-dimensional interference vector; $\mathbf{n}(i)$ is $n_o$-dimensional vector of noise; $\mathbf{H}_s(i)$ is $n_o\times m_o$-dimensional matrix; $\mathbf{H}_i(i)$ is a matrix of dimension $n_o\times k_o$; $\mathbf{x}_s(i)$, $\mathbf{x}_i(i)$ and $\mathbf{n}(i)$ are normally distributed independent random vectors (their mean values equal zero).

The covariance matrices of the observed process take the form:

$\eta_0$: $\mathbf{K}_0=[\mathbf{K}_0(i,j)]$; $\mathbf{K}_0(i,j)=E[\mathbf{H}_i(i)\mathbf{x}_i(i)\mathbf{x}_i^T(j)\mathbf{H}_i^T(j)]=$

$+E[\mathbf{n}(i)\mathbf{n}^T(j)]=\mathbf{H}_i(i)\mathbf{R}_i(i,j)\mathbf{H}_i^T(j)+\mathbf{N}(i)\delta_{ij}=$

$=\mathbf{K}_i(i,j)+\mathbf{N}(i)\delta_{ij}$; (2.1.2)

$\eta_1$: $\mathbf{K}_1=[\mathbf{K}_1(i,j)]$; $\mathbf{K}_1(i,j)=\mathbf{H}_s(i)\mathbf{R}_s(i,j)\mathbf{H}_s^T(j)+\mathbf{K}_0(i,j)=$

$=\mathbf{K}_{si}(i,j)+\mathbf{N}(i)\delta_{ij}=\mathbf{K}_s(i,j)+\mathbf{K}_i(i,j)+\mathbf{N}(i)\delta_{ij}$

where $\delta_{ij}$ is the Kronecker symbol; E[...] is a symbol of statistical average; superscript T means operation of transponating; $\mathbf{R}_s(i,j)=E[\mathbf{x}_s(i)\mathbf{x}_s^T(j)]$; $\mathbf{R}_i(i,j)=E[\mathbf{x}_i(i)\mathbf{x}_i^T(j)]$; $\mathbf{K}_0$ and $\mathbf{K}_1$ are block symmetrical matrices with the number of blocks in each equal $n\times n$ and with the number of elements in each block $\mathbf{K}_0(i,j)$ and $\mathbf{K}_1(i,j)$ equal $n_0\times n_0$.

Signal and interference can be presented as a Markov vector discrete random process. They are solutions of the following difference equations:

$$\mathbf{x}_s(i)=\mathbf{S}_s(i,i-1)\mathbf{x}_s(i-1)+\mathbf{G}_s(i-1)\mathbf{w}_s(i-1);$$
$$\mathbf{x}_i(i)=\mathbf{S}_i(i,i-1)\mathbf{x}_i(i-1)+\mathbf{G}_i(i-1)\mathbf{w}_i(i-1) \qquad (2.1.3)$$

where $\mathbf{S}_s(i,i-1)$ and $\mathbf{S}_i(i,i-1)$ are matrices of dimension $m_0\times m_0$ and $k_0\times k_0$; $\mathbf{G}_s(i-1)$ and $\mathbf{G}_i(i-1)$ are matrices of dimension $m_0\times l_0$ and $k_0\times q_0$; $\mathbf{w}_s(i-1)$ and $\mathbf{w}_i(i-1)$ are $l_0$- and $q_0$-dimensional vectors of uncorrelated mutually independent normal discrete random process; $E[\mathbf{w}_s(i)]=E[\mathbf{w}_i(i)]=\mathbf{0}$; $E[\mathbf{n}(i)\mathbf{w}_s^T(j)]=E[\mathbf{n}(i)\mathbf{w}_i^T(j)]=0$; $E[\mathbf{x}(i)\mathbf{w}_s^T(j)]=E[\mathbf{x}(i)\mathbf{w}_i^T(j)]=0$; $i<j$; $i,j=\overline{1,n}$.

The problem is to find the signal processing algorithm providing the best discrimination between the hypotheses $\eta_0$ and $\eta_1$ taking account of (2.1.3).

## 2.2. SYNTHESIS.

Let us show that the solution of the problem formulated reduces conceptually to the solution of the problem of the signals

$\mathbf{X}(i)=\mathbf{H}_s(i)\mathbf{x}_s(i)+\mathbf{H}_i(i)\mathbf{x}_i(i)$

and

$\mathbf{Y}(i)=\mathbf{H}_i(i)\mathbf{x}_i(i)$

detection against the background of uncorrelated in time normally distributed noises $\mathbf{n}(i)$.

It is known that a sufficient statistics, on the basis of which the problem stated is solved, [1-4] is

$$y_o(n)=\sum_{i=1}^{n}\sum_{j=1}^{n}\mathbf{z}^T(i)\mathbf{W}(i,j)\mathbf{z}(j) \qquad (2.2.1)$$

where $\mathbf{W}(i,j)=\mathbf{M}(i,j)-\mathbf{L}(i,j)$ (the matrices $\mathbf{M}(i,j)$ and $\mathbf{L}(i,j)$ are $n_o\times m_o$-blocks of inverse covariance matrices $\mathbf{K}_o^{-1}=[\mathbf{M}(i,j)]$ and $\mathbf{K}_1^{-1}=[\mathbf{L}(i,j)]$.

The matrix

$\mathbf{W}(i,j)=\mathbf{W}_X(i,j)-\mathbf{W}_Y(i,j)$

where $\mathbf{W}_X(i,j)=\mathbf{N}^{-1}(i)\delta_{ij}-\mathbf{L}(i,j)$, $\mathbf{W}_Y(i,j)=\mathbf{N}^{-1}(i)\delta_{ij}-\mathbf{M}(i,j)$ and the relationship (2.2.1) regarding this acquires the form

$$y_o(n)=y_X(n)-y_Y(n) \qquad (2.2.2)$$

where

$$y_X(n)=\sum_{i=1}^{n}\sum_{j=1}^{n}\mathbf{z}^T(i)\mathbf{W}_X(i,j)\mathbf{z}(j); \qquad (2.2.3)$$

$$y_Y(n)=\sum_{i=1}^{n}\sum_{j=1}^{n}\mathbf{z}^T(i)\mathbf{W}_Y(i,j)\mathbf{z}(j). \qquad (2.2.4)$$

The relationships (2.2.3) and (2.2.4) express a sufficient statistics for problem of signals $\mathbf{X}(i)$ (2.2.3) and $\mathbf{Y}(i)$ (2.2.4) detection against the background of uncorrelated noises $\mathbf{n}(i)$.

So the problem of the signal $\mathbf{H}_s(i)\mathbf{x}_s(i)$ detection on the background of mixture of time-correlated interferences $\mathbf{H}_i(i)\mathbf{x}_i(i)$ and uncorrelated noise $\mathbf{n}(i)$ reduces to the problem of stochastic signals $\mathbf{X}(i)$ and $\mathbf{Y}(i)$ detection on the background of uncorrelated noise $\mathbf{n}(i)$. Recursive relations, which solve this problem, are synthesized in the Section 1.1. Leaning upon these results we present the algorithms for the signal $\mathbf{H}_s(i)\mathbf{x}_s(i)$ detection on the background of correlated interferences and noises $\mathbf{H}_i(i)\mathbf{x}_i(i)+\mathbf{n}(i)$.

In accordance with the Scetion 1.1, the two variants of such algorithms are plausible.

### 2.2.1. **The first algorithm.**

The first of the algorithms consist in formation of difference between estimates $y_{11}(l)$ and $y_{10}(l)$ of sufficient statistics $y_X(l)$ and $y_Y(l)$, l is current time, $l=\overline{1,n}$:

$$y_1(l)=y_{11}(l)-y_{10}(l). \qquad (2.2.5)$$

The processes $y_{11}(l)$ and $y_{10}(l)$ are the output effects of the system for the signals $\mathbf{X}(i)$ and $\mathbf{Y}(i)$ detection on the background of noise $\mathbf{n}(i)$. They are farmed in accordance with recursive relations

$$y_{11}(l)=y_{11}(l-1)+\mathbf{z}^T(l)\mathbf{U}_{11}(l);\ y_{11}(0)=0; \qquad (2.2.6)$$

$$\mathbf{U}_{11}(l)=\mathbf{F}_1(l,l-1)\mathbf{U}_{11}(l-1)+\mathbf{W}_1(l)\mathbf{z}(l);\mathbf{U}_{11}(0)=\mathbf{0}; \qquad (2.2.7)$$

$$y_{10}(l)=y_{10}(l-1)+\mathbf{z}^T(l)\mathbf{U}_{10}(l);\ y_{10}(0)=0; \qquad (2.2.8)$$

$$\mathbf{U}_{10}(l)=\mathbf{F}_0(l,l-1)\mathbf{U}_{10}(l-1)+\mathbf{W}_0(l)\mathbf{z}(l);\mathbf{U}_{10}(0)=\mathbf{0}; \qquad (2.2.9)$$

The decision on validity of one of the hypotheses ($\eta_0$ or $\eta_1$)

is made as a result of comparison of $y_1(n)$ (or, depending on a detection criterion used, $y_1(l)$) with the predetermined threshold (or thresholds).

For the practical purposes to lower the number of operations used it is worthwhile fulfilling the realization of the algorithm (2.2.5) to (2.2.9) in some other way, and namely, the output signals $\mathbf{U}_{11}(l)$ and $\mathbf{U}_{10}(l)$ should come from the blocks realizing linear recurrent procedures (2.2.7) and (2.2.9) to substractor, where the difference is formed:

$$\mathbf{U}_1(l)=\mathbf{U}_{11}(l)-\mathbf{U}_{10}(l). \qquad (2.2.10)$$

Then nonlinear processing of the observed signal should be done:

$$y_1(l)=y_1(l-1)+\mathbf{z}^T(l)\mathbf{U}_1(l);\ y_1(0)=0; \qquad (2.2.6)$$

Thus the first variant of the algorithm, which solve the problem of the signal $\mathbf{H}_s(i)\mathbf{x}_s(i)$ detection against the background of the interference $\mathbf{H}_i(i)\mathbf{x}_i(i)+\mathbf{n}(i)$, is expressed by the relations (2.2.7), (2.2.9), (2.2.10) and (2.2.11).

The structural scheme of the obtained algorithm linear part, which realizes (2.2.7) and (2.2.9), is presented in Fig.2.2.1,a, the full structural scheme of the algorithm (2.2.7), (2.2.9):(2.2.11) is given in Fig.2.2.1,b. The block $\gamma_{del}=1$ means one-time-delay device, '1' is a shift register, '2' is a threshold device. The matrix parameters $\mathbf{W}_1(l)$ and $\mathbf{W}_0(l)$ have a role of amplification coefficients of processing system linear tracts, and the parameters $\mathbf{F}_1(l,l-1)$ and $\mathbf{F}_0(l,l-1)$ are feedback coefficients.

Let us present the relations allowing to obtain these parameters. As it is shown in the Section 1.1 the matrix parameter

$\mathbf{F}_1(l,l-1)$ is determined by the following relation:

$$\mathbf{F}_1(l,l-1)=\mathbf{L}_1(l)\mathbf{S}_1(l,l-1)\mathbf{N}(l-1) \qquad (2.2.12)$$

where

$$\mathbf{S}_1(l,l-1)=\Big[\mathbf{H}_s(l)\mathbf{S}_s(l.l-1)\mathbf{R}_s(l-1.l-1)\mathbf{H}_s^T(l-1)+ \\ +\mathbf{H}_i(l)\mathbf{S}_i(l,l-1)\mathbf{R}_i(l-1,l-1)\mathbf{H}_i^T(l-1)\Big]\mathbf{K}_{si}^{-1}(l-1,l-1) \qquad (2.2.13)$$

is a matrix which enters into the difference equation for the process

$$\mathbf{X}(l)=\mathbf{S}_1(l,l-1)\mathbf{X}(l-1)+\mathbf{G}(l-1)\mathbf{w}(l-1). \qquad (2.2.14)$$

In (2.2.14)

$$\mathbf{G}(l-1)\mathbf{w}(l-1)=\mathbf{H}_s(l)\mathbf{G}_s(l-1)\mathbf{w}_s(l-1)+\mathbf{H}_i(l)\mathbf{G}_i(l-1)\mathbf{w}_i(l-1).$$

The derivation of (2.2.13) is given in [17].

The matrix parameters $\mathbf{W}_1(l)$ represent the matrix $\mathbf{W}_1(i,j)=\mathbf{W}_1^T(j,i)$ for $i=j=l$, and in turn

$$\mathbf{W}_1(i,j)=\mathbf{N}^{-1}(i)\delta_{ij}-\mathbf{L}_1(i,j). \qquad (2.2.15)$$

The matrix $\mathbf{L}_1(i,j)$ is calculated by solving the following matrix equation (see Section 1.1):

$$\sum_{j=1}^{l}\mathbf{L}_1(i,j)\mathbf{K}_1(j,m)=\mathbf{I}\delta_{im};\ i,m<l, \qquad (2.2.16)$$

where $\mathbf{I}$ is the unit matrix of dimension $n_O\times n_O$. Remark that $\mathbf{L}_1(l)=\mathbf{L}_1(l,l)$ in (2.2.12).

The matrix parameters from (2.2.9) are determined by the relations

$$\mathbf{F}_O(i,i-1)=\mathbf{L}_O(l)\mathbf{S}_O(l,l-1)\mathbf{N}(l-1); \qquad (2.2.17)$$

$$\mathbf{W}_O(l)=\mathbf{W}_O(l,l), \qquad (2.2.18)$$

with

$$\mathbf{S}_O(l,l-1)=\mathbf{H}_i(l)\mathbf{S}_i(l,l-1)\mathbf{R}_i(l-1,l-1) \\ \mathbf{H}_i^T(l-1)\mathbf{K}_i^{-1}(l-1,l-1) \qquad (2.2.19)$$

or, if the matrix $\mathbf{H}_i^{-1}(l-1)$ exist,

$$\mathbf{S}_o(l,l-1)=\mathbf{H}_i(l)\mathbf{S}_i(l,l-1)\mathbf{H}_i^{-1}(l-1), \qquad (2.2.20)$$

$$\mathbf{W}_o(i,j)=\mathbf{W}_o^T(j,i)=\mathbf{N}^{-1}(i)\delta_{ij}-\mathbf{L}_o(i,j); \qquad (2.2.21)$$

the matrix $\mathbf{L}_o(i,j)$ is calculated from the equation

$$\sum_{j=1}^{l} \mathbf{L}_o(i,j)\mathbf{K}_o(j,m)=\mathbf{I}\delta_{im};\ i,m<l. \qquad (2.2.22)$$

In (2.2.17) $\mathbf{L}_o(l)=\mathbf{L}_o(l,l)$. The relations (2.2.19):(2.2.21) are deduced in [17].

2.2.2. **The second algorithm.**

Dwell on the second variant of the detection algorithm. It may be expressed via recursive relations for output effect $y_2(l)$ of the detection system and output signals $\mathbf{U}_{21}(l)$ and $\mathbf{U}_{20}(l)$ of the algorithm linear parts.

$$y_2(l)=y_2(l-1)+\mathbf{U}_{21}^T(l)\mathbf{U}_{21}(l)-\mathbf{U}_{20}^T(l)\mathbf{U}_{20}(l);\ \mathbf{y}_2(0)=0; \qquad (2.2.23)$$

$$\mathbf{U}_{21}(l)=\mathbf{F}_1(l,l-1)\mathbf{U}_{21}(l-1)+\mathbf{b}_1(l)\mathbf{z}(l);\ \mathbf{U}_{21}(0)=\mathbf{0}; \qquad (2.2.24)$$

$$\mathbf{U}_{20}(l)=\mathbf{F}_o(l,l-1)\mathbf{U}_{20}(l-1)+\mathbf{b}_o(l)\mathbf{z}(l);\ \mathbf{U}_{20}(0)=\mathbf{0}. \qquad (2.2.25)$$

The structural schemes of linear part of the algorithm, which solves the problem of the signal $\mathbf{X}(i)$ (or $\mathbf{Y}(l)$) detection on the noise $\mathbf{n}(i)$ background, is shown in Fig.2.2.2,a. This scheme is involved as an element in general structural scheme of the algorithm (2.2.23):(2.2.25) for detection of the signal $\mathbf{H}_s(l)\mathbf{x}_s(l)$ against a background of interferences and noises $\mathbf{H}_i(l)\mathbf{x}_i(l)+\mathbf{n}(l)$, which is given in Fig.2.2.2,b. The matrix parameters $\mathbf{b}_1(l)$ and $\mathbf{b}_o(l)$, involved in (2.2.24) and (2.2.25) and playing the role of amplification coefficients of the algorithm linear parts, are determined from the formulae (1.1.12) and

(1.1.13):

$$\mathbf{W}_1(i,j)=\sum_{k=1}^{n} \mathbf{b}_1(i,k)\mathbf{b}_1(k,j);\ \ \mathbf{b}_1(i,k)=\mathbf{b}_1^T(k,i); \qquad (2.2.26)$$

$$\mathbf{W}_0(i,j)=\sum_{k=1}^{n} \mathbf{b}_0(i,k)\mathbf{b}_0(k,j);\ \ \mathbf{b}_0(i,k)=\mathbf{b}_0^T(k,i); \qquad (2.2.27)$$

### 2.2.3. **Summary.**

The problem of stochastic signal detection against the background of any combination of stochastic interferences and noises may be solved using the algorithms obtained. It is necessary to remark that the algorithms given above solve, properly speaking, a two-alternative problem of the signals **X**(i) and **Y**(i) discrimination. The problems of multialternative detection (or discrimination) and methodically close to them problems of filtering and estimation of signal parameters may be solved similarly on the basis of the algorithms obtained.

## 3. MULTIALTERNATIVE DETECTION.

The problem of multialternative detection (or discrimination) of stochastic signals is directly connected with the problem of recognition of the objects reflecting or radiating the signal: the objects are known to impart to the signal in some state space, e.g. in the received signals space or in the space of parameters determining a trajectory or a movement characteristic of the object, a certain "tint", which is characteristic only for the object of certain type or class. This allow these two problems to be connected.

The principles of multialternative detection are founded on the theory of two-alternative detection of stochastic signals which is sufficiently well developed and summarized in a series of monographs [1-4,6]. The systems of Gaussian stochastic signals detection synthesized on the basis of this theory might be subdivided into two main groups: the systems founded on using the sufficient statistics in the form of a quadratic form in the exponent sign of the expression for the likelihood ratio [1,2,6], and the systems of joint detection-measuring which solve the problem of detection by bringing into the detector structure (e.g. by means of Kalman filtering) the parameters of the input signal [4,15,18]. The systems pertaining to the first group as a rule require a great volume of a priori statistical information, are very complex in realization [4] and are not adapted for using in the phase spaces. The systems of the second group are described in terms of the state space and have a more compact structure. However their application for multialterna-tive detection is problematic, since the signal parameter filtering in the channels, which correspond to error alternatives, might result in additional inexactitude in formation of the likelihood ratio, which is the basis for the problem's solution.

The synthesis of the system for two-alternative detection of stochastic signals given in the space state has been carried out in the Sections 1.1-2.2. The algorithms synthesized are of recursive character, their structure is sufficiently simple in comparison with the well-known systems; they assure the solution of the detection problem in various phase spaces. Nevertheless the use of these ones for multialternative detection of stochastic signals calls for a certain modification of them. This section is devoted to such a modification of one of the above algorithms.

### 3.1. PROBLEM STATEMENT.

The $n_o$-dimensional observation vector z(i) given in discrete time i, $i=\overline{1,n}$, may be formed in accordance with M+1 hypotheses $\eta_\mu$, $\mu=\overline{0,M}$:

$$\eta_\mu: \quad \mathbf{z}(i)=\mathbf{H}_\mu(i)\mathbf{x}_\mu(i)+\mathbf{n}(i). \tag{3.1.1}$$

In equation (1) $\mathbf{x}_\mu(i)$ is a $m_\mu$-dimensional signal vector; $\mathbf{n}(i)$ is a $n_o$-dimensional interference vector; $\mathbf{H}_\mu(i)$ is a matrix of dimension $n_o\times m_\mu$; $\mathbf{x}_\mu(i)$ and $\mathbf{n}(i)$ are normally distributed, independent, zero-mean random vectors. The covariance matrices of the observed process $\mathbf{z}(i)$ are:

$$\begin{gathered}\eta_\mu: \ \mathbf{K}_\mu=[\mathbf{K}_\mu(i,j)]; \\ \mathbf{K}_\mu(i,j)=\mathbf{K}_{s\mu}(i,j)+\mathbf{K}_\mu(i,j)=\mathbf{H}_\mu(i)\mathbf{R}_{s\mu}(i,j)\mathbf{H}_\mu^T(j)+\mathbf{N}(i)\delta_{ij},\end{gathered} \tag{3.1.2}$$

where $\mathbf{R}_{s\mu}(i,j)=E[\mathbf{x}_\mu(i)\mathbf{x}_\mu^T(j)]$; $\delta_{ij}$ is the Kronecker symbol; $E[/]$ means a symbol of statistical average, the sign 'T' means transponation. The matrices $\mathbf{K}_\mu$ represents block square symmetrical ones with a number of $\mathbf{K}_\mu(i,j)$-blocks in each of them being $n\times n$ and with a number of elements in the blocks equal to $n_o\times n_o$. Let us assume that hypothesis $\eta_o$ corresponds to the signal absence at the system input: $\mathbf{x}_o(i)=\mathbf{0}$.

Each of the signals $\mathbf{x}_\mu(i)$, $\mu\neq 0$, can be given as a Markov vector random process and is a solution of the following linear difference equation:

$$\mathbf{x}_\mu(i)=\mathbf{S}_\mu(i,i-1)\mathbf{x}_\mu(i-1)+\mathbf{G}_\mu(i-1)\mathbf{w}_\mu(i-1), \qquad (3.1.3)$$

where $\mathbf{S}_\mu(i,i-1)$ is a matrix of dimension $m_\mu\times m_\mu$; $\mathbf{w}_\mu(i-1)$ is a $l_\mu$-dimensional vector of uncorrelated normal noise; $\mathbf{G}_\mu(i-1)$ is a matrix of dimension $m_\mu\times l_\mu$; $E[\mathbf{w}_\mu(i)]=\mathbf{0}$; $E[\mathbf{n}(i)\mathbf{w}_\mu^T(j)]=0$; $E[\mathbf{x}_\mu(i)\mathbf{w}_\mu^T(j)]=0$; $i<j$; $i,j=\overline{1,n}$; $\mu=\overline{1,M}$. It is necessary to find the signal processing algorithm which discriminates between the hypotheses $\eta_\mu$, $\mu=\overline{0,M}$, in the best way.

## 3.2. SYNTHESIS OF ALGORITHM FOR DISCRIMINATION OF SIGNALS.

The solution of the proposed problem leans upon the results of the synthesis of the recurrent algorithm for stochastic signals detection obtained in the Section 1.1.

Traditional methods for determining systems, which detect signals $\mathbf{x}_\mu(i)$ in the noise $\mathbf{n}(i)$, as a rule, are based on the quadratic form in the exponent sign of the expression for the likelihood ratio; this quadratic form

$$y_{\mu}(n)= \frac{1}{2}\sum_{i=1}^{n}\sum_{j=1}^{n} \mathbf{z}^{T}(i)\mathbf{W}_{o\mu}(i,j)\mathbf{z}(j) \qquad (3.2.1)$$

is a sufficient statistic for the problem of two-alternative detection. In the expression (3.2.1)

$$\mathbf{W}_{o\mu}(i,j)=\mathbf{N}^{-1}(i)\delta_{ij}-\mathbf{L}_{o\mu}(i,j).$$

The matrix $\mathbf{L}_{o\mu}(i,j)$ is determined from the solution of the following linear equations system:

$$\sum_{j=1}^{n} \mathbf{L}_{o\mu}(i,j)\mathbf{K}_{\mu}(j,m)= \mathbf{I}\ \delta_{ij};\ i,m<n; \qquad (3.2.2)$$

where $\mathbf{I}$ is the $n_o$-dimensional unit matrix.

The recurrent detection algorithm synthesized in the Section 1.1 is equivalent to the ratio (see (1.1.8)), the subscript $\mu$ of the corresponding functions is subsequently omitted:

$$y(n)= \sum_{l=1}^{n} \mathbf{z}^{T}(l)\mathbf{U}(l);\ \mathbf{U}(l)= \sum_{j=1}^{l} \mathbf{W}(l,j)\mathbf{z}(j); \qquad (3.2.3)$$

where $\mathbf{W}(i,j)=\mathbf{N}^{-1}(i)\delta_{ij}-\mathbf{L}(i,j)$; the matrix $\mathbf{L}(i,j)$ is a block-element of the "current" inverse matrix $\mathbf{K}_{\mu l}^{-1}$ obtained when the matrix $\mathbf{K}_{\mu}$ is inverted at the current-time interval $i,j=\overline{1,l}$, $l$ is the current time, $l<n$:

$$\sum_{j=1}^{l} \mathbf{L}(i,j)\mathbf{K}_{\mu}(j,m)= \mathbf{I}\delta_{im};\ i,m<l,\ l<n. \qquad (3.2.4)$$

Thus the algorithm (3.2.3) differs from the "traditional" one (3.2.1) by the character of the matrix $\mathbf{K}_{\mu}$ inversion and in that it allows for only the left lower part of matrix $\mathbf{W}=[\mathbf{W}(i,j)]$, $j<i$, $i<l$. As a result of mentioned peculiarities of the statistics of $y(n)$ we have succeeded in synthesizing the following rather compact recurrent detection algorithm (1.1.26):

$$
\begin{cases}
y(l)= y(l-1)+\mathbf{z}^T(l)\mathbf{U}(l);\ y(0)=0; \\
\mathbf{U}(l)= \mathbf{F}(l,l-1)\mathbf{U}(l-1)+\mathbf{W}(l)\mathbf{z}(l);\ \mathbf{U}(0)=\mathbf{0};\ l=\overline{1,n}.
\end{cases}
\tag{3.2.5}
$$

In (3.2.5) $\mathbf{W}(l)=\mathbf{W}(l,l)$; matrix $\mathbf{F}(l,l-1)$ is given by the expression:

$$\mathbf{F}(l,l-1)= \mathbf{L}(l)\mathbf{H}(l)\mathbf{S}(l,l-1)\mathbf{H}^{-1}(l-1)\mathbf{N}(l-1). \tag{3.2.6}$$

$\mathbf{H}^{-1}(l-1)$ exists; if $\mathbf{H}(l-1)$ is not a square matrix, $\mathbf{H}^{-1}(l-1)$ should be considered as quasi-inverse one; $\mathbf{L}(l)=\mathbf{L}(l,l)$.

Following the Section 1.1, the matrix $\mathbf{F}(l,l-1)$ connects $\mathbf{W}(l,m)$ and $\mathbf{W}(l-1,m)$ matrices obtained for the adjacent moments of current time:

$$\mathbf{W}(l,m)= \mathbf{F}(l,l-1)\mathbf{W}(l-1,m);\ m<l-1. \tag{3.2.7}$$

To solve the problem stated in the section it is necessary to pass from the statistics (3.2.3) to the logarithm of likelihood ratio obtained with regard for the peculiarities of $\mathbf{K}_\mu$ matrix inversion. This log-likelihood ratio is a sufficient statistic for the multialternative detection problems.

Let us modify the expression (3.2.3) to supplement the triangular matrix $\mathbf{W}=[\mathbf{W}(i,j)]$, $j<i$; $i,j<l$, with the terms $\mathbf{W}(j,i)=\mathbf{W}^T(i,j)$ to symmetrical one for imparting to the obtained expression the shape analogous to the quadratic form (3.2.1):

$$
\begin{aligned}
y_1(n) &= \frac{1}{2}\sum_{l=1}^{n} \mathbf{z}^T(l) \sum_{j=1}^{n} \mathbf{W}(l,j)\mathbf{z}(j)= \\
&= y(n) - \frac{1}{2}\sum_{l=1}^{n} \mathbf{z}^T(l)\mathbf{W}(l)\mathbf{z}(l).
\end{aligned}
\tag{3.2.8}
$$

The algorithm (3.2.5), which is conformable to $y_1(n)$, is:

$$\begin{cases} y_1(l)=y_1(l-1)+\Delta y_1(l); \\ \Delta y_1(l)=\mathbf{z}^T(l)[\mathbf{U}_1(l)-\frac{1}{2}\mathbf{W}(l)\mathbf{z}(l)]; \; y_1(0)=0; \\ \mathbf{U}_1(l)=\mathbf{F}(l,l-1)\mathbf{U}_1(l-1)+\mathbf{W}(l)\mathbf{z}(l); \; \mathbf{U}_1(0)=\mathbf{0}; \; l=\overline{1,n}. \end{cases} \quad (3.2.9)$$

Regarding the Gaussian character of the observed signals distribution, equations (3.2.9) are equivalent to the following expression for the log-likelihood ratio $\Lambda(L)$:

$$\Lambda(l)= \Lambda(L-1)+\alpha(l); \; \Lambda(0)=0; \; l=\overline{1,n}, \quad (3.2.10)$$

where

$$\alpha(l)=ln \frac{p[\mathbf{z}(l)|\mathbf{Z}_{l-1},\eta_\mu]}{p[\mathbf{z}(l)|\mathbf{Z}_{l-1},\eta_o]};$$

$p[\mathbf{z}(l)|\mathbf{Z}_{l-1},\eta_{\mu(o)}]$ is the conditional density of vector $\mathbf{z}(l)$ probability distribution, when the vector $\mathbf{Z}_{l-1}=\{\mathbf{z}(1),\mathbf{z}(2),\mathbf{z}(3),...,\mathbf{z}(l-1)\}$ is known, and the hypothesis $\eta_{\mu(o)}$, $\mu=\overline{1,M}$ is given; when l=1, the function $p[\mathbf{z}(1)|\eta_{\mu(o)}]$ corresponds to this probability density.

The expressions for the corresponding probability densities are as follows:

$$p[\mathbf{z}(l)|\mathbf{Z}_{l-1},\eta_\mu]= \sqrt{\frac{c(l)}{(2\pi)^{n_o}}} \, exp\{-\tfrac{1}{2}\mathbf{Z}_l^T\mathbf{A}_l\mathbf{Z}_l\}; \quad (3.2.11)$$

$$\mathbf{Z}_l^T=\{\mathbf{z}^T(1),\mathbf{z}^T(2),\mathbf{z}^T(3),...,\mathbf{z}^T(l)\};$$

$$p[\mathbf{z}(l)|\mathbf{Z}_{l-1},\eta_o]=$$

$$=p[\mathbf{z}(l)|\eta_o]= \frac{1}{\sqrt{(2\pi)^{n_o}|det\mathbf{N}(l)|}}\times \quad (3.2.12)$$

$$\times exp\{-\tfrac{1}{2}\mathbf{z}^T(l)\mathbf{N}^{-1}(l)\mathbf{z}(l)\};$$

$$p[\mathbf{z}(l)|\eta_\mu]=\frac{1}{\sqrt{(2\pi)^{n_o}|det\mathbf{K}_\mu(l,l)|}}\,exp\{-\tfrac{1}{2}\mathbf{z}^T(l)\mathbf{L}(l)\mathbf{z}(l)\};$$

$$p[\mathbf{z}(l)|\eta_o]=\frac{1}{\sqrt{(2\pi)^{n_o}|det\mathbf{N}(l)|}}\,exp\{-\tfrac{1}{2}\mathbf{z}^T(l)\mathbf{N}^{-1}(l)\mathbf{z}(l)\},$$

where

$$\mathbf{A}_l = \begin{bmatrix} \begin{pmatrix} \; \end{pmatrix} & & \mathbf{L}^T(l,1) \\ & & \mathbf{L}^T(l,2) \\ & & \vdots \\ \mathbf{L}(l,1) & \mathbf{L}(l,2)\ldots & \mathbf{L}(l) \end{bmatrix};$$

$$c(l) = \left|\frac{det\mathbf{L}_l}{det\mathbf{L}_{l-1}}\right|;$$

$$\mathbf{L}_l = \begin{bmatrix} \mathbf{L}(1) & \mathbf{L}^T(2,1)\ldots & \mathbf{L}^T(l,1) \\ \mathbf{L}(2,1) & \mathbf{L}(2) & \ldots\mathbf{L}^T(l,2) \\ \vdots & \vdots & \vdots \\ \mathbf{L}(l,1) & \mathbf{L}(l,2) & \ldots\mathbf{L}(l) \end{bmatrix}.$$

In the designations of the submatrices $\mathbf{L}(k,j)$ in the expression for $\mathbf{L}_l$ the first argument corresponds to the value of current time within the limits of which the inversion of matrix (3.2.4) is made; when the time correlation of the process $\mathbf{x}(l)$ is sufficiently weak or when the signal-to-interference ratio is small $c(l)\approx|det\mathbf{L}(l)|$. Under the above conditions, the matrix $\mathbf{L}^{-1}(l)$ might be considered as that of vector $\mathbf{z}(l)$ conditional covariances when given vector $\mathbf{Z}_{l-1}$ and the hypothesis $\eta_\mu$ is valid.

Taking into the account equations (3.2.11) and (3.2.12), the value $\alpha(l)$ is determined as

$$\alpha(l)=\tfrac{1}{2}[ln|det\mathbf{N}(l)|-ln|c(l)|]-\Delta y_1(l).$$

The value $\Delta y_1(l)$ (3.2.9) may be expressed as

$$\Delta y_1(l)=\frac{1}{2}\mathbf{Z}_1^T\mathbf{B}_1\mathbf{Z}_1,$$

where

$$\mathbf{B}_1 = \begin{bmatrix} & \bigcirc & -\mathbf{L}^T(l,1) \\ & & -\mathbf{L}^T(l,2) \\ & & \vdots \\ -\mathbf{L}(l,1) & -\mathbf{L}(l,2)\ldots & [\mathbf{N}^{-1}(l)-\mathbf{L}(l)] \end{bmatrix} =$$

$$= \begin{bmatrix} & \bigcirc & \mathbf{W}^T(l,1) \\ & & \mathbf{W}^T(l,2) \\ & & \vdots \\ \mathbf{W}(l,1) & \mathbf{W}(l,2)\ldots & \mathbf{W}(l) \end{bmatrix}. \qquad (3.2.13)$$

Taking account of (3.2.7), the matrices $\mathbf{W}(l,k)$ constituting (3.2.13) are connected with matrices $\mathbf{W}(k)$, $k=\overline{1,l-1}$, as follows:

$$\mathbf{W}(l,k)= \mathbf{F}(l,l-1)\mathbf{F}(l-1,l-2)\ldots\mathbf{F}(k+1,k)\mathbf{W}(k).$$

The recurrent procedure of the log-likelihood ratio calculation may be given in the final shape by the expression (the subscript $\mu$ is written again):

$$\begin{cases} \Lambda_\mu(l)=\Lambda_\mu(l-1)+\alpha_\mu(l); \; \Lambda_\mu(0)=0; \; l=1,n; \\ \alpha_\mu(l)=d_\mu(l)+\Delta y_{1\mu}(l); \\ d_\mu(l)=\frac{1}{2}[\mathit{ln}|\mathit{det}\mathbf{N}(l)|-\mathit{ln}|c(l)|]; \\ \Delta y_{1\mu}(l)=\mathbf{z}^T(l)[\mathbf{U}_{1\mu}(l)-\frac{1}{2}\mathbf{W}_\mu(l)\mathbf{z}(l)]; \\ \mathbf{U}_{1\mu}(l)=\mathbf{F}_\mu(l,l-1)\mathbf{U}_{1\mu}(l-1)+\mathbf{W}_\mu(l)z(l); \\ \mathbf{U}_{1\mu}(0)=0; \; l=\overline{1,n}; \; \mu=\overline{1,M}. \end{cases} \qquad (3.2.14)$$

The structural scheme of the algorithm (3.2.14) is shown in Fig.3.2.1. The decision on the signal presence or absence is made by comparing the $\Lambda_\mu(n)$ quantity with the given threshold. The block $\gamma_{del}=1$ in Fig.3.2.1 means the one-time-delay device.

In order to solve the problem of signal discrimination, M algorithmic channels (3.2.14) generating the logarithm of the likelihood ratio $\Lambda_{\mu}(n)$ for each of M signals are created; when $\Lambda_{\mu}(n) \geqslant h$ ( h is a detection threshold ) and $\Lambda_{\mu}(n) > \Lambda_{\gamma}(n)$ for all $\gamma \neq \mu$ the decision is made on $\eta_{\mu}$; if $\Lambda_{\mu}(n) < h$ for all $\mu$ the decision is made that only the noise is present ( the hypothesis $\eta_{o}$).

All the parameters and matrix coefficients in (3.2.14) are calculated in advance.

### 3.2.1. **Recursive procedure for matrix $\mathbf{L}(l)$ calculation.**

Realization of the above algorithm for multialternative detection of stochastic signals requires the knowledge at every step of observation the elements $\mathbf{L}(l)$ of the matrix $\mathbf{K}_{\mu l}^{-1}$, which is the inverse one with respect to the left upper submatrices $\mathbf{K}_{\mu l}$ (with the number of blocks $l \times l$) of the matrix $\mathbf{K}_{\mu}$ (the subscript $\mu$ of the matrix $\mathbf{L}(l)$ is rejected). The matrix $\mathbf{L}(l)$ is used for calculation of $\mathbf{F}(l,l-1)$ and $\mathbf{W}(l)$ parameters of the above algorithm. For calculation of this matrix the procedure based on inversion of block matrices [19] may be proposed, which allows to obtain the matrix $\mathbf{K}_{\mu l}$ from the results of inversion of the matrix $\mathbf{K}_{\mu l-1}$ with the number of blocks $(l-1) \times (l-1)$. This procedure is described below.

Let us rewrite the matrix $\mathbf{K}_{\mu l}$ in the form

$$\mathbf{K}_{\mu l} = \left[ \begin{array}{c:c} \mathbf{K}_{\mu l-1} & \mathbf{A}(l) \\ \hdashline \mathbf{A}^{T}(l) & \mathbf{K}_{\mu}(l) \end{array} \right] \qquad (3.2.15)$$

where $\mathbf{K}_{\mu}(l) = \mathbf{K}_{\mu}(l,l)$; $\mathbf{A}(l)$ is a block matrix-column composed of $(l-1)$ blocks.

The matrix inverse with respect to $\mathbf{K}_{\mu l}$ equals

$$K^{-1}_{\mu l} = \left[\begin{array}{c|c} \mathbf{B}(l) & \mathbf{C}(l) \\ \hline \mathbf{C}^{T}(l) & \mathbf{L}(l) \end{array}\right] \tag{3.2.16}$$

where the dimension of the introduced submatrices corresponds to the dimension of those in (3.2.15). Using the above designations, the algorithm for calculation of the matrix $\mathbf{K}^{-1}_{\mu l-1}$ (and consequently the matrix $\mathbf{L}(l)$), if the matrix $\mathbf{K}_{\mu l-1}$ is known, is adduced as follows: since l=2

$$\begin{cases} \mathbf{L}(l)= \mathbf{D}^{-1}_{l}; \\ \mathbf{D}_{l}= \mathbf{K}_{\mu}(l)- \mathbf{A}^{T}(l)\mathbf{K}^{-1}_{\mu}(l); \\ \mathbf{B}(l)= \mathbf{K}^{-1}_{\mu l-1}+ \mathbf{K}^{-1}_{\mu l-1}\mathbf{A}(l)\mathbf{D}^{-1}_{l}\mathbf{A}^{T}(l)\mathbf{K}^{-1}_{\mu l-1}; \\ \mathbf{C}(l)= - \mathbf{K}^{-1}_{\mu l-1}\mathbf{A}(l)\mathbf{D}^{-1}_{l}. \end{cases} \tag{3.2.17}$$

**3.4. SUMMARY.**

The modification of one of the detection algorithms synthesized in the Section 1.1 is made, directed to substitution for sufficient statistics on which the algorithms have been founded. While for two-alternative detection of Gaussian stochastic signals the known quadratic form with a kernel of a difference between two inverse correlation matrices corresponding to the alternatives considered being the sufficient statistics, for the problems of the signals discrimination a logarithm of the likelihood ratio is the sufficient statistics. The algorithm modified in such a way forms the logarithm of the likelihood ratio for every of possible hypothesis by use of the recurrent procedure, keeping all the advantages of the algorithms synthesized formerly. Some details may be seen in [20,21].

## CONCLUSION.

The problem of synthesis of recursive algorithms for stochastic signal detection in phase space, which has been stated in the paper, has been solved. The algorithms synthesized and considered in this paper allow to detect and discriminate (in case of multialternative detection) stochastic signals in various phase spaces. Their structure is rather simple, they are properly the analogs of a Kalman filter in the detection theory, this allow unifying system elements either for detection or for posterior detected signal filtering. The detection efficiency of the algorithms synthesized practically does not differ from one of the well-known optimal stochastic signal detection algorithm (1.2.2). Their feedback and amplification coefficients are stabilized at the first steps of observation. Recursive character of the algorithms significantly reduces the required computational powers and make them convenient for use in digital systems. Additional details concerning the above algorithms may be seen in [22].

## TRANSLATOR'S ACKNOWLEDGMENTS.

The translator would like to express his sincerest gratitude to all the colleges who helped him in preparation of this paper and especially to Dr. V.V.Pivovarov of the Institute of Space Recearches of the Russian Academy of Sciences whose numerical calculations are depicted in the Section 1.2.2.

**FIGURE CAPTIONS.**

Fig. 1.1.1. Recurrent detection algorithms: (a) algorithm (1.1.26); (b) algorithm (1.1.27); (c) the Kalman filter; $\gamma_{del} = 1$ means one-time-delay device; the linear parts are in the dashed boxes.

Fig. 1.2.1. Dynamics of amplification W(l) and feedback F(l, l – 1) coefficients of the algorithm (1.1.26).

Fig. 1.2.2. The results of computations of logarithm of the inverse detection probability as a function of signal-to-noise ratio $q = p^{-1}$ of the system realizing the algorithm (1.1.26); the formula (1.2.5) was used.

Fig. 1.2.3. The directivity characteristics of the space-time processing system realizing the algorithm (1.1.26); 1 – $\sigma^2_{n0} = 10$, $\sigma^2_{n1} = 0$; 2 – $\sigma^2_{n0} = 10$, $\sigma^2_{n1} = 10$; 3 – $\sigma^2_{n0} = 1$, $\sigma^2_{n1} = 10$.

Fig. 1.3.1. The synthesized filters (a, b) and the continuous Kalman filter (c): (a) (1.3.20) and (1.3.19); (b) (1.3.21) and (1.3.18); (c) the Kalman filter.

Fig. 1.3.2. Dependences of amplification (W) and feedback ($F_0$) factors of the detection filter (1.3.20), (1.3.19) on signal-to-interference-correlation-time ratio δ.

Fig. 2.2.1. A structural scheme of the recurrent algorithm (2.2.7), (2.2.9): (2.2.11): (a) linear part; (b) complete scheme. 1 – shift register; 2 – threshold device; $\gamma_{del} = 1$ is a one-time-delay device.

Fig. 2.2.2. A structural scheme of the recurrent algorithm (2.2.23):(2.2.25): (a) linear part; (b) complete scheme. 1 – shift register; 2 – threshold device; $\gamma_{del} = 1$ means a one-time-delay device.

Fig. 3.2.1. Recurrent algorithm for multialternative detection of stochastic signals in the state space: 1 – comparator; 2 – decision (is made when $l = n$); 3 – from other channels; $\gamma_{del} = 1$ is a one-time-delay device.

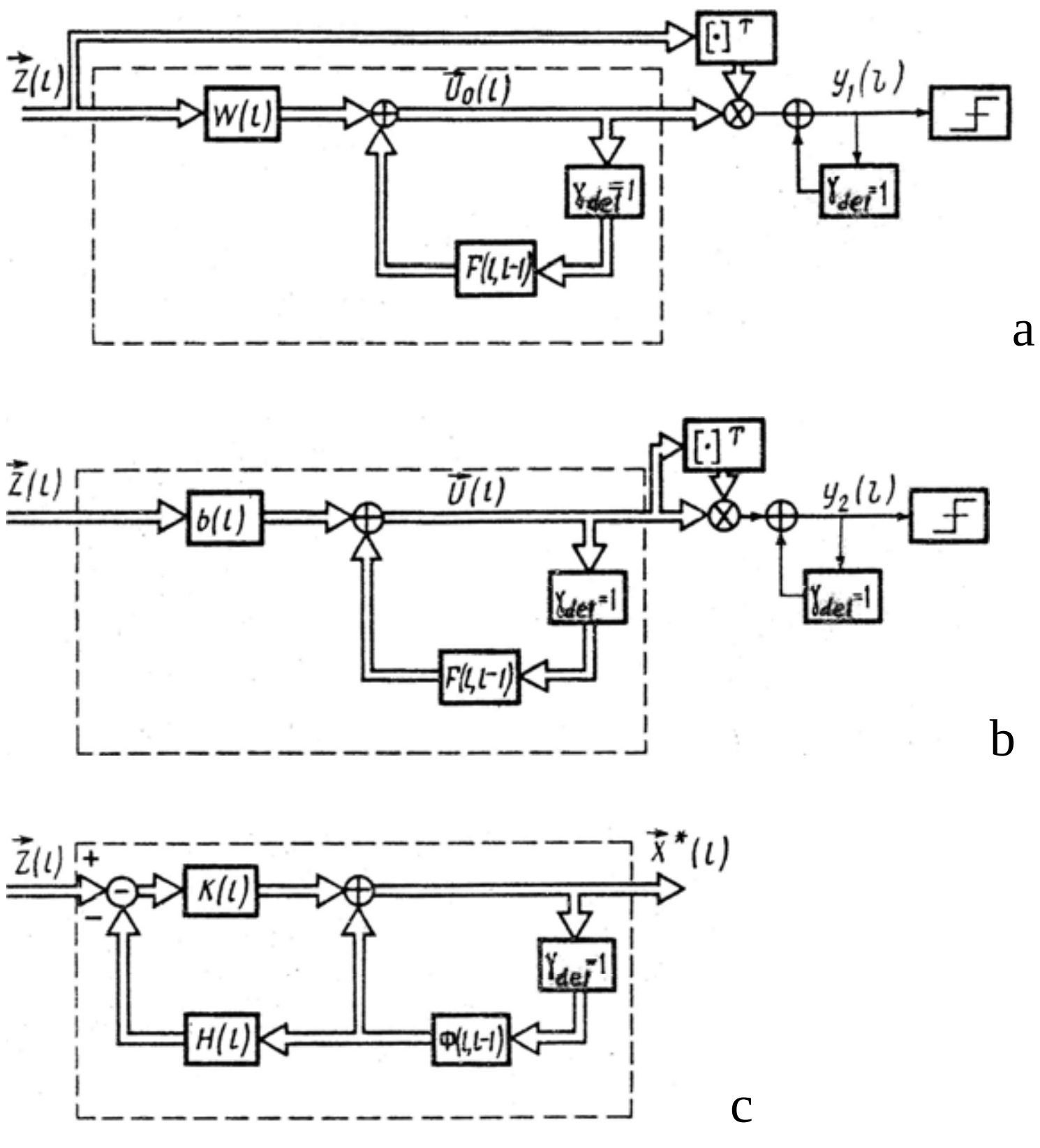

Fig. 1.1.1.

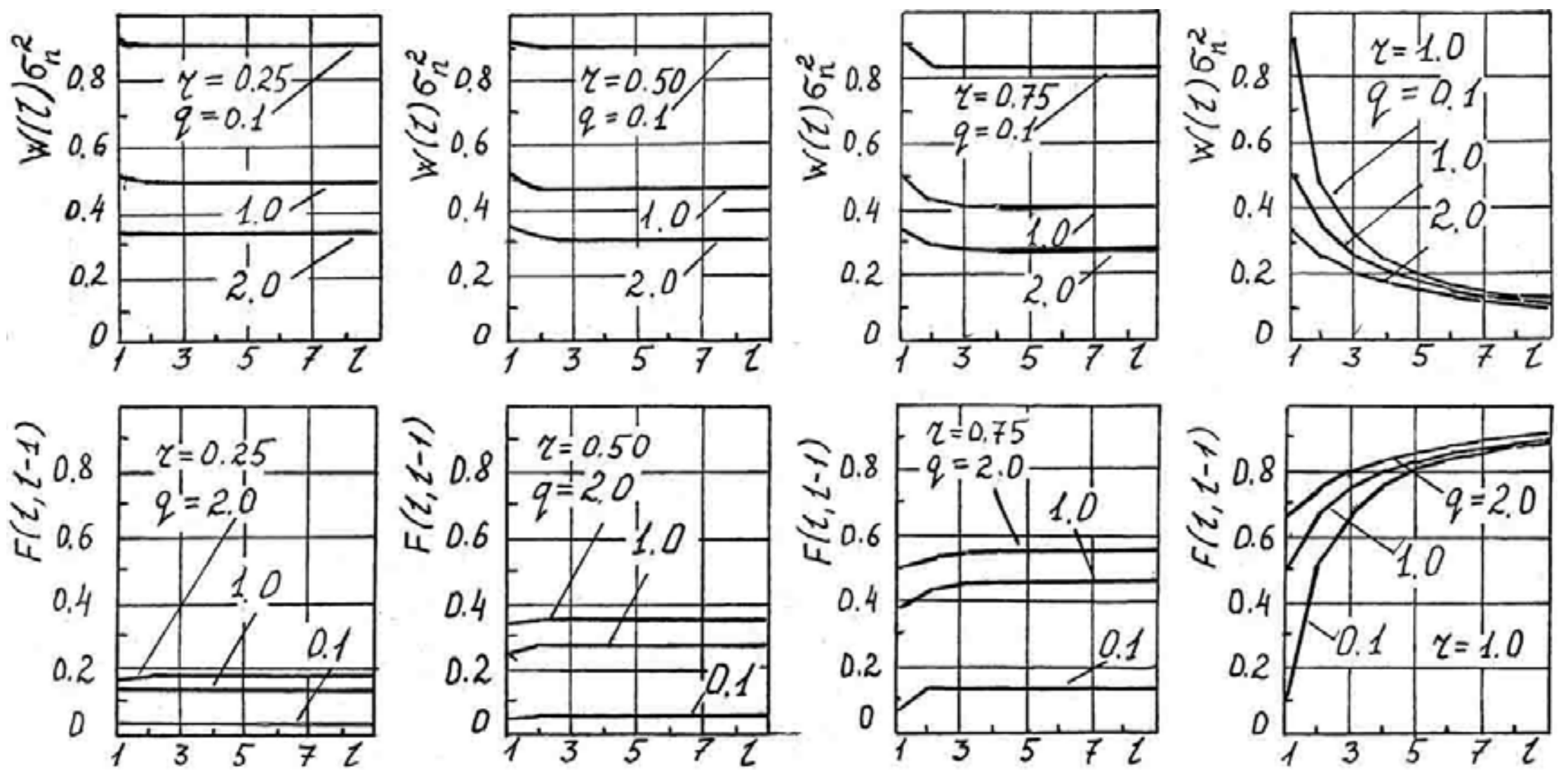

Fig. 1.2.1.

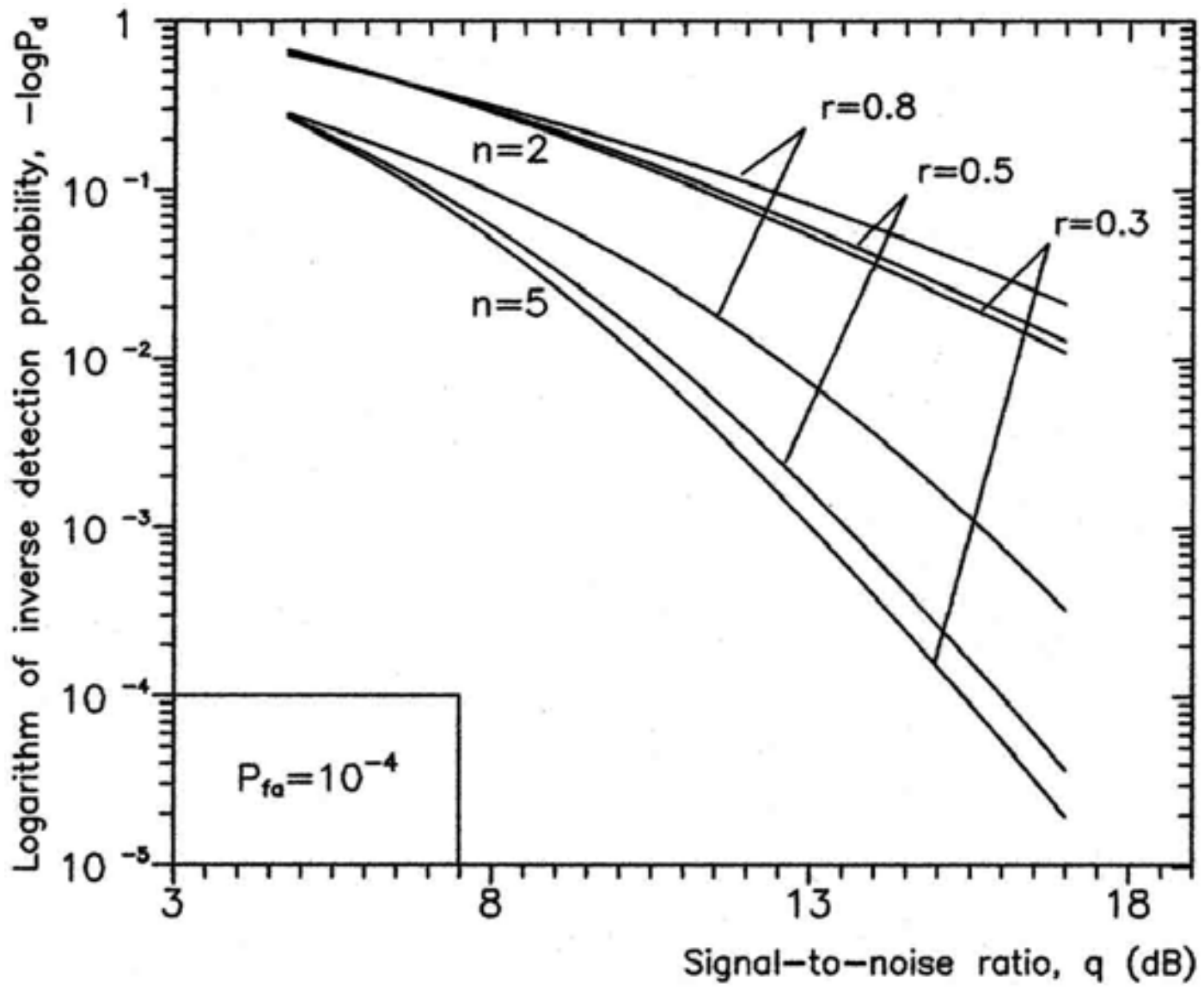

Fig. 1.2.2.

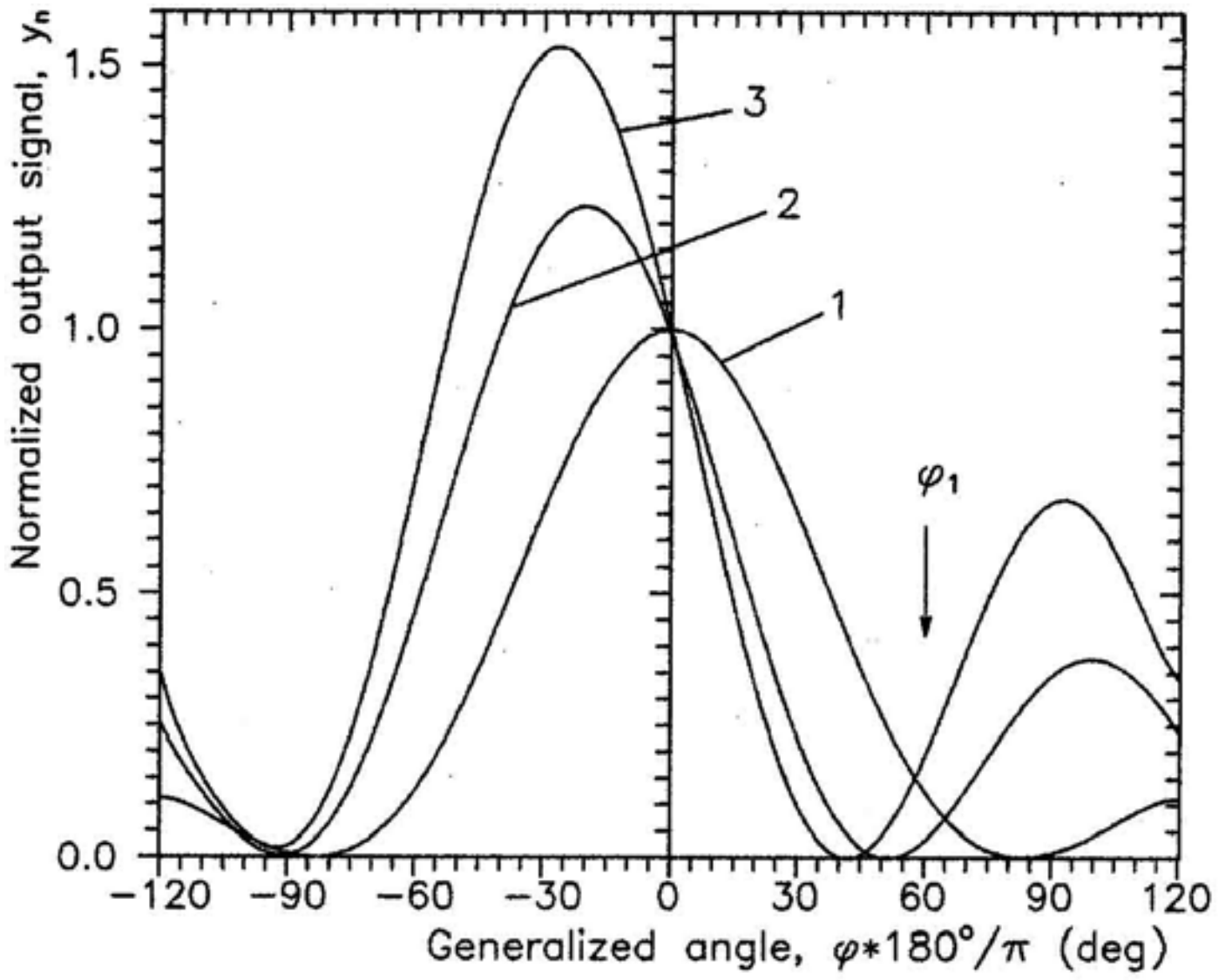

Fig. 1.2.3.

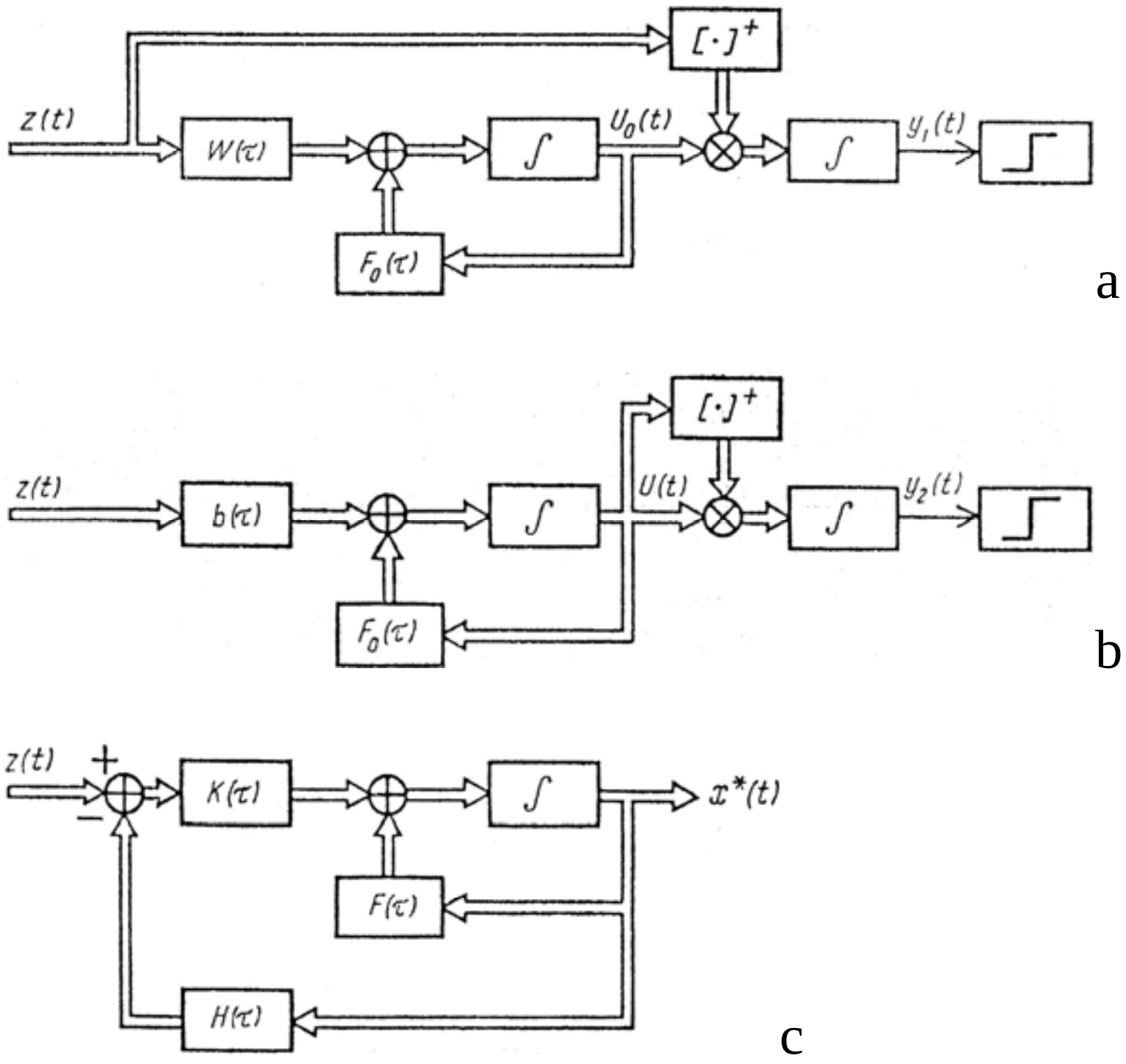

Fig. 1.3.1.

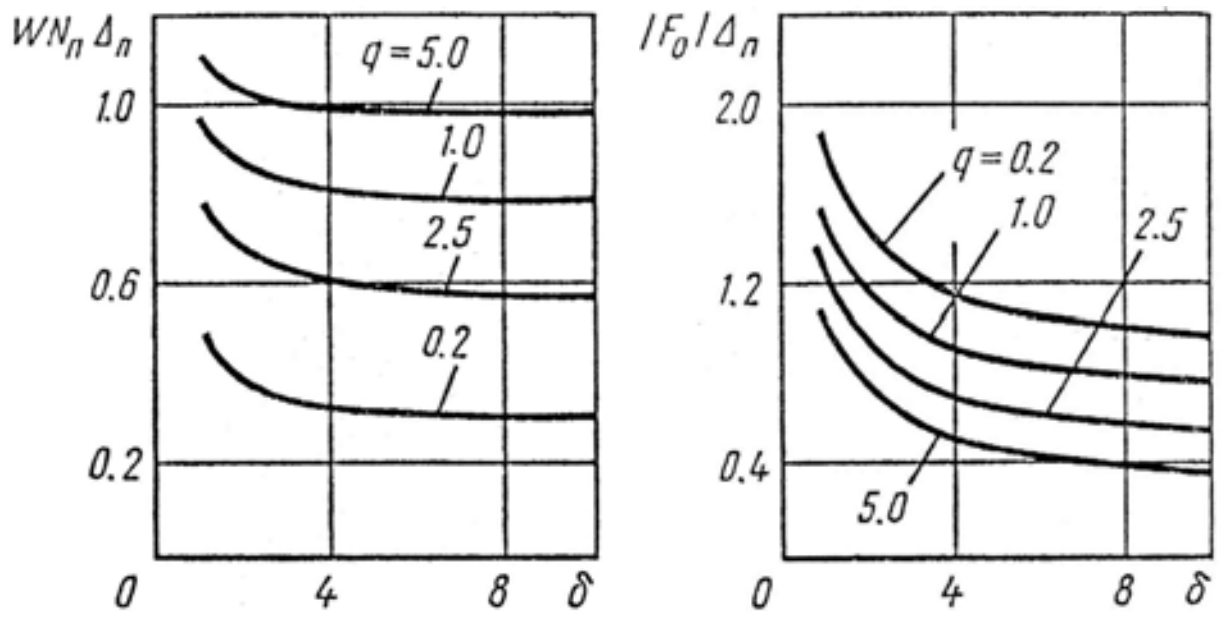

Fig. 1.3.2.

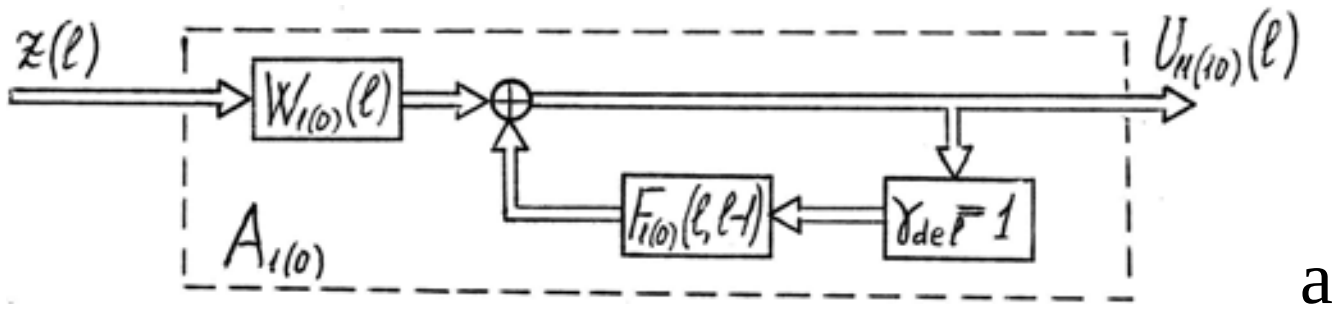

a

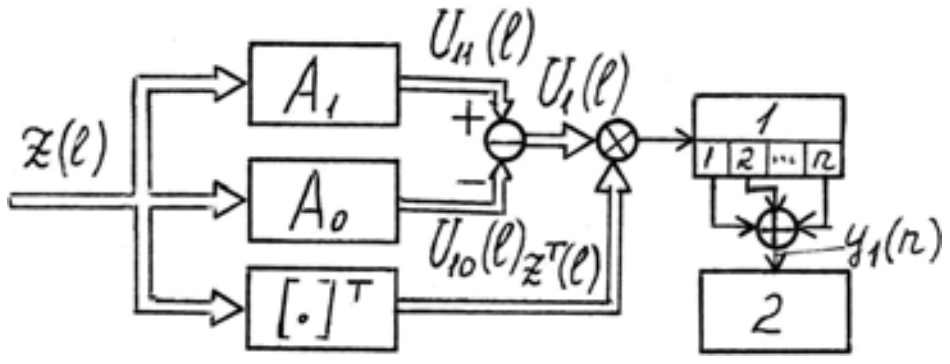

b

Fig. 2.2.1.

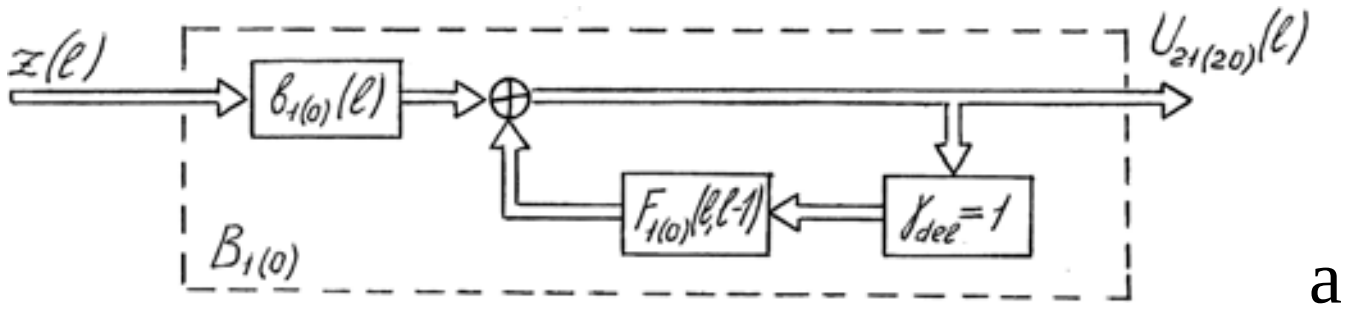

a

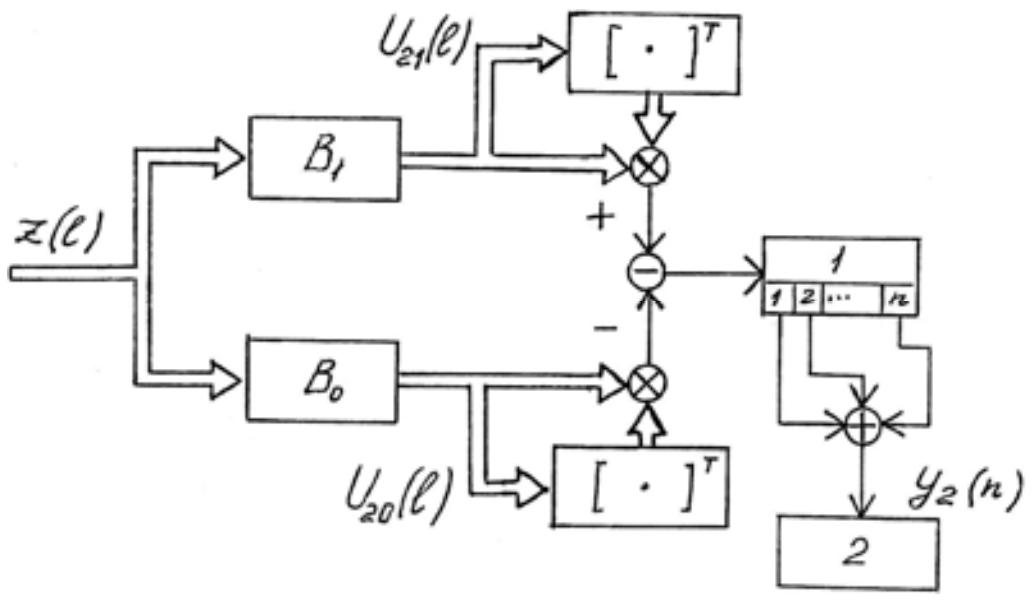

b

Fig. 2.2.2.

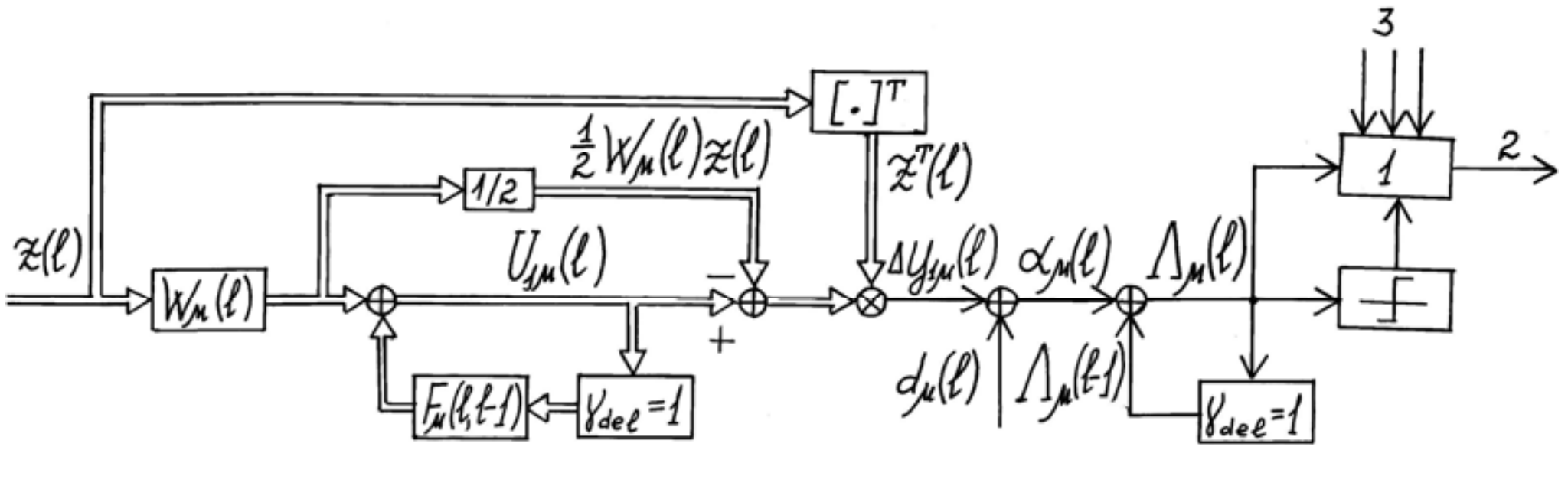

Fig. 3.2.1.